\documentclass[twocolumn]{emulateapj}
\pdfoutput=1 
\usepackage{graphicx}
\usepackage{natbib}
\usepackage{multirow}
\usepackage{textcomp}
\usepackage{longtable}

\newcommand{\bc}{\begin{center}}
\newcommand{\ec}{\end{center}}

\newcommand{\hii}{HII }
\newcommand{\am}{NH$_{3}$}
\newcommand{\cyano}{HC$_3$N}

\newcommand{\cm}{cm$^{-3}$}
\newcommand{\meth}{CH$_3$OH}
\newcommand{\methcy}{CH$_3$CN}

\newcommand{\msun}{M$_{\odot}$}

\newcommand{\kms}{km s$^{-1}}

\newcommand{\jyb}{Jy beam$^{-1}}
\newcommand{\h}{$^{\mathrm{h}}$}
\newcommand{\m}{$^{\mathrm{m}}$}
\newcommand{\s}{$^{\mathrm{s}}}
\newcommand{\amm}{NH$_{3}}
\newcommand{\thebrick}{GCM0.253+0.016}
\newcommand{\bricktwo}{M0.25+0.01}
\newcommand{\brickthree}{G0.253+0.016}
\newcommand{\carc}{``C$-$arc''}
\newcommand{\bricks}{the Brick}

\begin{document}

\slugcomment{}

\title{Abundant \meth\,  Masers but no New Evidence for Star Formation in \thebrick}

\author{E.A.C. Mills\altaffilmark{1}}
\affil{National Radio Astronomy Observatory\altaffilmark{2} 1003 Lopezville Rd Socorro, NM 87801}
\email{bmills@aoc.nrao.edu}

\author{N. Butterfield, D.A. Ludovici}
\affil{Department of Physics and Astronomy, University of Iowa, Iowa City, IA 52245}

\author{C.C. Lang\altaffilmark{3}$^{,}$\altaffilmark{4}}
\affil{\altaffilmark{3}Department of Physics and Astronomy, University of Iowa, Iowa City, IA 52245}
\affil{\altaffilmark{4}School of Mathematics and Physics, University of Tasmania Private Bag 37, Hobart, Tasmania 7001 Australia}

\author{J. Ott}
\affil{National Radio Astronomy Observatory\altaffilmark{2} 1003 Lopezville Rd Socorro, NM 87801}

\author{M.R. Morris} 
\affil{Department of Physics and Astronomy, University of California, 430 Portola Plaza, Box 951547 Los Angeles, CA 90095-1547}

\author{S. Schmitz}
\affil{Department of Physics and Astronomy, University of Iowa, Iowa City, IA 52245}

\altaffiltext{1}{E.A.C. Mills is a Jansky Fellow of the National Radio Astronomy Observatory.}
\altaffiltext{2}{The National Radio Astronomy Observatory is a facility of the National Science Foundation operated under cooperative agreement by Associated Universities, Inc.}

\begin{abstract}
We present new observations of the quiescent giant molecular cloud \thebrick~in the Galactic center, using the upgraded Karl G. Jansky Very Large Array. 
Observations were made at wavelengths near 1 cm, at K (24 to 26 GHz) and Ka (27 and 36 GHz) bands, with velocity resolutions of 1$-$3 \kms$ and spatial resolutions of $\sim$ 0.1 pc, at the assumed 8.4 kpc distance of this cloud. The continuum observations of this cloud are the most sensitive yet made, and reveal previously undetected emission which we attribute primarily to free-free emission from external ionization of the cloud. In addition to the sensitive continuum map, we produce maps of 12 molecular lines: 8 transitions of \am--(1,1),(2,2),(3,3),(4,4),(5,5),(6,6),(7,7) and (9,9), as well as the \cyano~(3$-$2) and (4$-$3) lines, and \meth\, $4_{-1}-3_0$ the latter of which is known to be a collisionally-excited maser. We identify 148  \meth\, $4_{-1}-3_0$ (36.2 GHz) sources, of which 68 have brightness temperatures in excess of the highest temperature measured for this cloud (400 K) and can be confirmed to be masers. The majority of these masers are concentrated in the southernmost part of the cloud. We find that neither these masers nor the continuum emission in this cloud provide strong evidence for ongoing star formation in excess of that previously inferred by the presence of an H$_2$O maser. 
\end{abstract}


\section{Introduction}
Molecular gas in the central 500 parsecs of the Galaxy (the Central Molecular Zone or CMZ) is concentrated in a population of giant molecular clouds with sizes of 15 to 50 pc, and masses of 10$^4-10^6$ \msun\, \citep[e.g.,][]{Molinari11}. On scales of a few parsecs, these CMZ clouds are characterized by large, turbulent linewidths, \citep[15$-$50~\kms$\hspace{-0.2cm},][]{Bally87}, high gas temperatures \citep[50$-$300~K,][]{Huttem93b, Mauers86}, and substantial densities \citep[$n > 10^{4}$~cm$^{-3}$,][]{Zylka92}. However, apart from the Sgr B2 cloud \citep[in which there are dozens of compact and hypercompact \hii regions as well as two massive hot cores and numerous water masers, indicating an extremely active star-forming environment,][]{Cheung69b,Vogel87,Gaume90,dePree98}, most CMZ clouds show little evidence of recent or ongoing star formation \citep{GD83,Ho85,Morris89,Morris93,Lis94,Caswell96,Immer12}.  

Exactly why CMZ clouds exhibit so little ongoing star formation is unclear. Given that the total amount of molecular gas in this region \citep[$\sim3\times10^7$ \msun,][]{Dahmen98} is just under 5\% of the total molecular gas in the Galaxy \citep[$\sim8.4\times10^8$ \msun,][]{Nakanishi06}, and the star formation rate in the CMZ makes up a similar fraction of the total estimated star formation rate in the Galaxy \citep{Longmore13}, there would not immediately appear to be a discrepancy. The difference is that gas in the CMZ is believed to be on average two orders of magnitude more dense than elsewhere in the Galaxy and might thus be expected to be forming stars at a proportionately higher rate \citep{Lada12,Longmore13}. If the CMZ deviates from relations between the amount of dense (n$>10^4$~cm$^{-3}$) gas and star formation which hold in other galaxies, this could suggest that star formation might proceed differently in such extreme environments \citep[e.g.,][]{Kruijssen14b}. Or, it may indicate that the ongoing star formation in CMZ clouds is underestimated using traditional indicators. It has also been suggested that we may be observing many CMZ clouds at a special time, just before the (possibly triggered) onset of star formation \citep{Longmore13b,Kruijssen15}. A final consideration is that, especially in regions with short orbital timescales like the CMZ, one must be careful to compare the amount of gas and star formation on spatial scales sufficiently large (and timescales sufficiently long) for them to be correlated \citep{Kruijssen14a}. Ultimately, whether or not the star formation process in the CMZ is truly unusual, observing CMZ clouds lacking in star formation is a unique opportunity to investigate the initial conditions of (massive) star formation in an extreme environment, before the star formation process itself begins to affect and further disrupt that environment. 

\subsection{\thebrick}
\thebrick~(also, \brickthree, G0.216+0.016, \bricktwo, M0.25+0.11,  or ``The Brick'', as it has been variously referred to in the literature) is one such extremely quiescent CMZ cloud, located $\sim$45 pc in projection from the dynamical center of the Galaxy \citep[assuming a Galactocentric distance of 8.4 kpc;][]{Ghez08, Gillessen09, Reid14}. Although its appearance as a prominent infrared dark cloud indicates that it is occulting most of the infrared emission from the nuclear bulge; its chemistry, kinematics, high temperatures, and large linewidths are all consistent with being located at the CMZ, and it is commonly taken to lie on the near side of the CMZ  \citep{Lis98}.  In total, \thebrick~ is suggested to have a mass of $1-2\times10^5$ \msun, making it one of the five most massive clouds in the CMZ \citep{Lis94,Longmore12}. It is also the only compact  $>10^4$ solar mass cloud found thus far in the entire Galaxy which does not exhibit advanced stages of star formation \citep{Ginsburg12,Tackenberg12,Urquhart14}. The comparably massive Maddalena cloud in the outer Galaxy, which also does not show evidence of active massive star formation, is extended over $\sim$ 100 pc. \citep{Madd85,Megeath09}. The large mass and relatively high average density of this cloud \citep[$n \sim 1\times10^5$ \cm;][]{Longmore12,Kauffmann13} suggest that it is capable of massive star and perhaps even cluster formation \citep{Longmore12}. However, there is no clear evidence in this cloud for ongoing massive star formation apart from a single water maser \citep{Lis94}.

Recent continuum studies of \thebrick\, at infrared to radio wavelengths have continued to search for signposts of ongoing star formation. \cite{Longmore12} analyze Herschel observations of the cloud and find no embedded heating sources at wavelengths up to 70 $\mu$m. At 280 GHz with the SMA, \cite{Kauffmann13} find only one strong, compact dust core, which they suggest is indicative of a low potential for star formation \citep[but see also][who find more extensive dust continuum emission at 230 GHz, also with the SMA]{Johnston14}. Both \cite{Johnston14} and \cite{Rathborne14b} then measure the column density probability distribution function from the dust continuum, finding it to be log-normal. \cite{Rathborne14b} do find a deviation from this form at high column densities (interpreted as self-gravitation), but state that this corresponds to just the single dust core already known to contain a water maser. The only potential indications of more advanced star formation come from high-resolution radio observations by \cite{Rodr13} who identify three compact thermal continuum sources which they suggest could be embedded B-stars. However, all of these sources are located outside of the bulk of the gas and dust emission in the cloud, on its periphery.

Although continuum observations show few signs of previously-missed star formation and are largely consistent with \thebrick~being in a quiescent, pre-star forming stage, a  host of recent observations of the gas reveal many other complexities. Velocity dispersions on spatial scales of 0.07-0.1 pc are observed to range from extremely turbulent \citep[$>30 $\kms$ as measured in ALMA observations of SO;][]{Higuchi14} to clumps with line widths apparently less than 1 \kms$ \citep[from SMA observations of the quiescent gas tracer N$_2$H$^+$, ][]{Kauffmann13}-- the narrowest line widths yet observed in a CMZ cloud. Abundant emission from other shock-tracing molecules SiO and \meth\, are additionally observed in \thebrick~by \cite{Rathborne15} and \cite{Johnston14}. Johnston et al. also present the first resolved temperature measurements of the cloud, using H$_2$CO, which indicate extremely high temperatures in the clumpy gas (T$\sim$300 K), much higher than temperatures measured from single-dish observations of the same lines \citep{Ao13}, though comparable to temperatures measured in this cloud from single-dish observations of highly-excited lines of \am\, \citep{Mills13}. The most surprising new observations are of a series of HCO$^+$ absorption filaments, suggested to be tracing the surface magnetic field lines in the cloud, which which have never before been seen in any giant molecular cloud \citep{Bally14}.

Complementing this existing suite of molecular line observations, we present the first interferometric study of both the molecular and ionized gas in \thebrick~at radio wavelengths, using the Karl G. Jansky Very Large Array (hereafter, `VLA'), a facility of the National Radio Astronomy Observatory\footnotemark[1]\footnotetext[1]{The National Radio Astronomy Observatory is a facility of the National Science Foundation operated under cooperative agreement by Associated Universities, Inc.}. The upgraded VLA offers more sensitive receivers and broad spectral bandwidths up to 8 GHz for both sensitive radio continuum imaging and spectral line surveys. Our new VLA observations of \thebrick~exploit both of these capabilities to provide a comprehensive new study of the sub-parsec morphology, kinematics and physical conditions of gas in this cloud. Our radio continuum observations allow for a more sensitive search for signs of ongoing star formation to determine whether indications of high-mass star formation in CMZ clouds may have been previously missed. With our molecular line data, we can examine kinematics of both low and high density gas in the cloud, using \am, an abundant tracer of gas having densities greater than a few $10^3$ \cm. The \am\, observations are also sensitive to gas over a wide range of temperatures (from tens to hundreds of K), enabling us to investigate the full range of temperatures present in the pre-star forming gas, and to map the temperature structure of that gas across the cloud. In this paper, we first present an analysis of the continuum emission and the morphology of the detected molecular species in \thebrick, with detailed analyses of the kinematic and temperature structure to be presented in subsequent papers.

In Section \ref{obs}, we describe the VLA observations and the procedures used to calibrate and image these data. We then present an overview of our study of \thebrick, beginning with the properties of the continuum emission which are analyzed in Section \ref{res}. The morphology and kinematics of the molecular gas are subsequently analyzed in Section \ref{morph}.  In Section \ref{mas}, we focus on emission from the 36.2 GHz \meth\, line and present a catalog of more than 70 candidate collisionally-excited masers. Finally, in Section \ref{dis}, we discuss constraints on the amount and nature of ongoing star formation in this cloud. 

\section{Observations and Data Calibration}
\label{obs}

The observations presented in this paper were made with the new WIDAR correlator of the VLA. The data were taken in two different frequency bands on two separate dates: in Ka-band (27$-$36 GHz) on January 13th, 2012, and in K-band (24$-$25 GHz) on January 14th, 2012, under project code 11B-210.  Both observations used the hybrid DnC array configuration to compensate for the low altitude of the Galactic center as observed from the VLA site. In this paper we present only the observation of \thebrick, however these data are part of a larger survey of the radio continuum and molecular line emission in CMZ clouds which will be described further in additional papers. 

\subsection{Observation Setup} 
\label{setup}

These observations employed the WIDAR correlator on the VLA in order to simultaneously observe both a wide spectral bandwidth for continuum studies and a large number of spectral lines. Observations in each band (K and Ka) are divided into two separate, continuous subbands of $\sim$0.84 GHz width which are each subdivided into 7 spectral windows.  In K-band, the subbands were centered on 24.054 GHz and 25.375 GHz, and for Ka-band the subbands were centered on 27.515 GHz and 36.35 GHz. 
The typical spectral resolution per spectral window is 250 kHz, with 512 channels per spectral window. However, for three spectral windows, covering (1) the \am\, (1,1) and (2,2) lines and their hyperfine satellites, (2) the 36.6 GHz \meth\, maser line, and (3) the \methcy\, K=2$-$1 transitions, the resolution was increased (125 kHz, $\sim$ 1-1.5 \kms$) to better resolve the line structure. 
 
In order to map the majority of \thebrick~in the K and Ka bands we used two pointings (see Table \ref{Sources}) oriented along the major axis of the cloud, which is elongated in declination. The total integration time for each field was $\sim$25 minutes. The distance between the pointings was $83.5\arcsec$ at K band and $70.7\arcsec$ at Ka band.  
Primary beam sizes range from $\sim110\arcsec$ at K band to $\sim100\arcsec$ and $\sim75\arcsec$ for the lower and upper frequencies observed at Ka-band, respectively. 

\subsection{Calibration}

The calibration of our K and Ka band VLA observations was performed using the CASA software provided by NRAO\footnotemark[2]\footnotetext[2]{http://casa.nrao.edu/}. 3C286 was observed as the flux density calibrator, J1733-1304 was used as the bandpass calibrator, and J1744-3116 was used as the gain and phase calibrator. Reference pointing was performed every hour on the phase calibrator, which was observed every $\sim$ 25 minutes throughout the observations.  In addition, at these higher VLA frequencies, corrections were made for the atmospheric opacity, which is determined from a mix of both actual weather data and a seasonal model using the {\it plotweather} task.  

\subsection{Imaging}
 
The K and Ka band continuum and spectral data were imaged using the CLEAN task in CASA.  As stated in Section \ref{setup}, two pointings were observed to cover the cloud at both K and Ka bands, which were mosaicked together when imaging. Continuum data were obtained by flagging out the spectral lines and end channels. Briggs weighting with a robust parameter of 0.5 in order to balance the point-source resolution with the sensitivity, giving synthesized beams of $1.59\arcsec$ to $2.30\arcsec$. The rms noise levels in the continuum images ranges from 29 to 55 $\mu$\jyb$. These values are generally about twice the theoretical rms noise levels of 17 to 27 $\mu$\jyb$, which may be due to the increased contributing emission coming from the Galactic plane. The largest angular scale to which the data are sensitive is $\sim60\arcsec$ at K band (or $\sim$2.4 pc) and $\sim40\arcsec$ ($\sim$1.6 pc) at the higher-frequency Ka-band subband. 

The K and Ka band spectral line data were also imaged using CLEAN. All lines were imaged individually, and continuum emission was subtracted using the CASA task {\it imcontsub}. The data were imaged at their intrinsic frequency resolution with no smoothing, and all spectral lines (except for the 36 GHz \meth\, line) were imaged with natural weighting, resulting in synthesized beams which ranged from 1.58 to 1.77$\arcsec$. For the 36 GHz \meth\, line, which exhibited primarily point-source emission, the data were imaged using Briggs weighting with a robust parameter of 0.5. Beam parameters for all images are given in Table \ref{Imaging}. 

For the \amm$ (1,1)$- $(6,6) and \cyano~(3$-$2) \& (4$-$3) transitions we used multiscale deconvolution to maximize sensitivity to extended emission in the cloud. The images were cleaned using beams with sizes of 0, 5, 20, and 80 pixels, and the parameter setting the relative weighting between these scales (0$-$80) was set to 0.6.

In the resulting multiscale images, `negative bowl' features indicative of missing extended flux were minimized, however the observations are still not sensitive to emission on scales larger than $\sim1\arcmin$. We did not use multiscale CLEAN for the \amm$ (7,7)$- $(9,9) and \methcy~(2$-$1) lines, as it was not found to significantly improve the imaging of these weak lines, or for the 36 GHz \meth\, line, for which the emission was primarily unresolved or on small spatial scales. The individual imaging parameters for each line can be found in Table \ref{Imaging}. In general, the rms noise levels per channel range from 0.7 to 3.0 m\jyb$ with the exception of \meth~(4$-$3). The typical rms levels for this transition varied, from 2.16 m\jyb$ in maser-free channels to 116 in the brightest maser channel. These rms values are consistent with the theoretical rms values of 1.4 m\jyb$ and 0.9 m\jyb$ for channel widths of 0.125 MHz and 0.250 MHz, respectively.

\subsection{Self-Calibration}  

For observations of the Ka band (36 GHz) \meth\, line, in which there are many strong point sources, we additionally self-calibrated the data. For each pointing, a bright \meth\, point source with minimal additional emission surrounding the source was chosen for self-calibration. Here, the requirement that the point sources be relatively isolated was more important than that they be the strongest in the cube.  Each pointing was first imaged using CLEAN with a small number of iterations, thus producing a map and model of the emission.  This model was used by the CASA task gaincal for both phase and amplitude calibrations. To begin with, phase-only self calibration was applied until the signal-to-noise improvement was no longer significant (2-3 iterations). After phase-only calibration, a single iteration of amplitude and phase self-calibration was performed. The self-calibration amplitude and phase solutions for this single channel were then applied to all of the channels in each pointing, and the pointings were  jointly imaged using CLEAN as described above to form a final image.

\section{Continuum Results and Analysis}
\label{res}

\begin{figure*}
\includegraphics[scale=0.8]{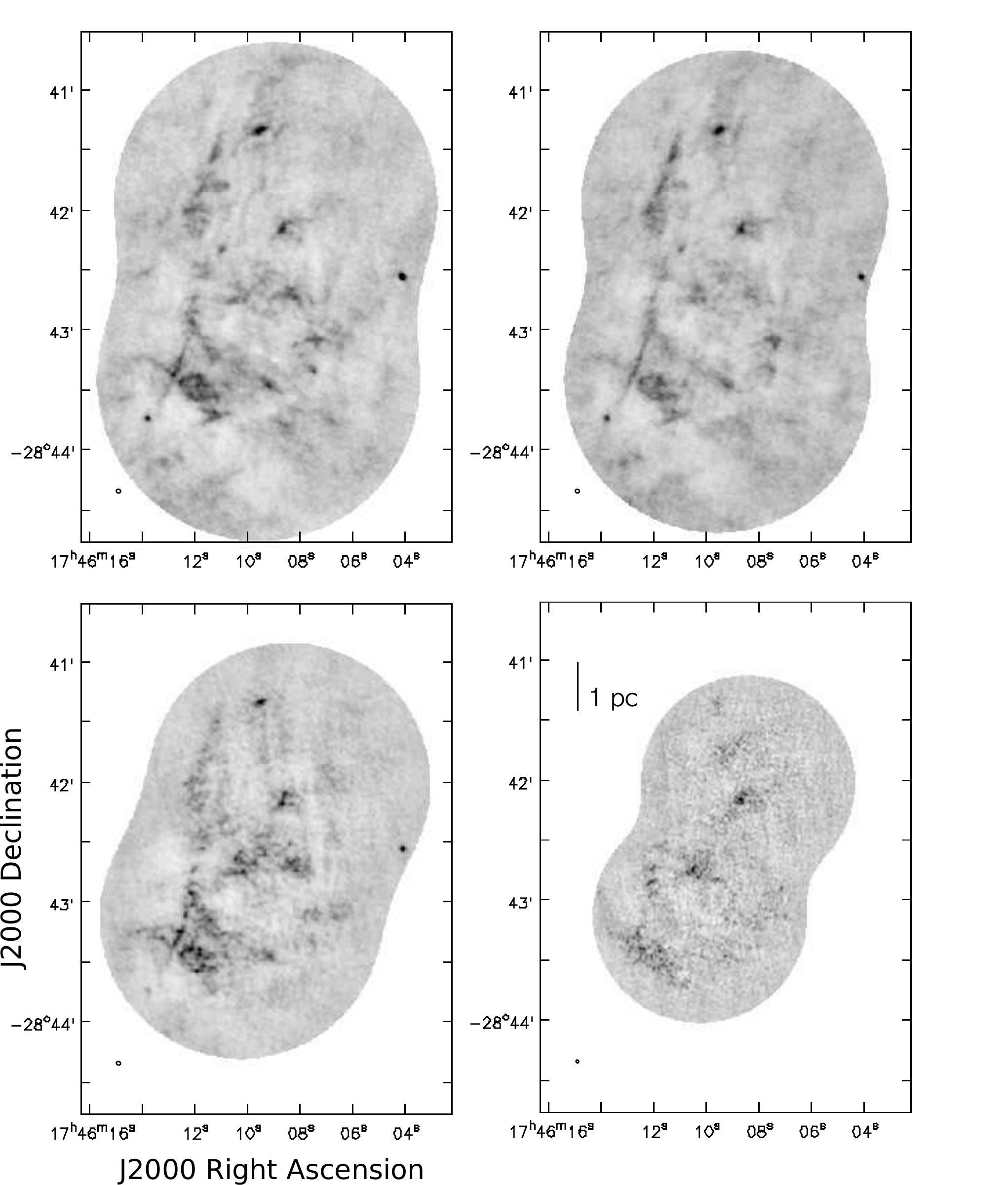}
\caption{ Continuum images at the four observation frequencies: 24.11 GHz, 25.43 GHz, 27.45 GHz, and 36.4 GHz. All four sub-images were cleaned using the same restoring beam: $2.295\arcsec\times1.966\arcsec$, PA=70.8075\degr, which corresponds to the beam size of the lowest frequency. The RMS noise of the image at each of the four frequencies is 89.3, 63.1, 120, and 80.3 $\mu$\jyb$ respectively.}
\label{4panel}
\end{figure*}

Figure \ref{4panel} shows the continuum emission associated with \thebrick~at 24.1 GHz (upper left), 25.4 GHz (upper right), 27.5 GHz (lower left) and 36.4 GHz (lower right). Table \ref{Imaging} presents the properties of each of these continuum images, including the parameters of the synthesized beam and the rms noise level. 

The radio continuum emission in the field of \thebrick~is fairly weak and typically extended over the same region in which gas and dust emission from the cloud has been detected \citep{Kauffmann13,Johnston14,Rathborne14b}. Exceptions to this are several compact and brighter sources which, as previously noted by \cite{Rodr13}, are generally located outside of the majority of gas and dust emission associated with \thebrick. Apart from these sources, the good coincidence in the spatial distribution and (as will be discussed further below), the morphology of the radio continuum and molecular gas emission, makes it likely that this continuum emission is truly associated with the \thebrick~cloud, and does not just arise from other sources along this confused line of sight. As the continuum images have not been corrected for missing emission at large spatial scales with the addition of single dish data, some of the continuum emission may be resolved out, especially at the highest frequencies (36.4 GHz). At these frequencies, significant flux may be missing from structures larger than $\sim$2$-$2.5 pc, affecting the calculations of spectral indices for these structures. The low-level, diffuse continuum emission, where it is detectable (mostly at 24.1 and 25.4 GHz) has a typical intensity of $\sim$0.2 m\jyb$, with a few brighter clumps that extend up to $\sim 0.7$ m\jyb$. In the following section, we identify 10 regions that are a good representation of the continuum emission associated with this cloud and analyze the possible nature and origin of this emission.

\subsection{Continuum Morphologies}

The regions of continuum emission that we will evaluate here are shown in Figure 2. The regions are defined by a contour levels of 6 or 10 times the 24.1 GHz rms noise level of 30 $\mu$Jy beam$^{-1}$)m depending on whether the regions are extended or compact, respectively. Table 3 summarizes the properties of these regions of continuum emission. In addition to presenting flux measurements for these regions from each of the VLA continuum images, we give the 90 GHz fluxes from both ALMA 12m-only images and single-dish-corrected ALMA images from \cite{Rathborne14b}.

Out of the 10 selected areas of interest, 4 regions (C2, C7, C8 and C9) are large and diffuse ($>$30\arcsec/1 pc across) and are primarily located in the Eastern part of the field. Of these, C9, located in the southern part of the field, is the largest and brightest at 24.1 GHz. Contained within the 6$\sigma$ contours of this region is a large elliptical or ``shell''-like region having a long axis extent of $\sim30\arcsec$ as well as a long thin ``filament'' of emission running from northwest to southeast that lies tangential to the elliptical region, with the two intersecting at $\alpha$(J2000)=$17^{\mathrm{h}}$46$^{\mathrm{m}}$12.7$^{\mathrm{s}}$, $\delta$(2000)=$-28\degr43\arcmin24\arcsec$. This tangential filament is $\sim$35\arcsec~in length, corresponding to a physical size of 1.4 pc. Although some clumpy emission in the region of the shell and filament is detected above the noise in the 36.4 GHz image, the size of the region is on the order of the largest angular size at this frequency, and likely its emission has been suppressed. Both shell and filament are still faintly seen at 3 mm \citep{Rathborne14b}, though their structure is not apparent at 230.9 GHz \citep{Johnston14}. 

Regions C7 and C8 have very similar structures and together form an apparent curved ridge of continuum emission across the cloud. While these two regions appear to fall along the same structure, they are separate at the 6$\sigma$ level, and we treat them as separate regions. We note that, like C9, the large-scale diffuse structure of these sources has likely been suppressed, and if there were a true connection between them, the present data would not be sensitive to emission on that spatial scale. At 36.4 GHz, where the suppression of large-scale emission is most severe, only the brightest knots in these sources are detected above the noise at this frequency. The morphologies of these two regions are also reminiscent of that of the adjacent molecular gas in the clouds, as we will discuss further in Section \ref{cmorph}. 

Like C9, the continuum emission in the most northern diffuse region, C2, exhibits filamentary structure. The eastern edge of C2 forms a straight line that is parallel to the tangential filament in C9. The straight edge of C2 also extends northward to an additional clump that is separate in the 6$\sigma$ contour levels. The total length of this linear feature, including the additional northern clump of emission, is $\sim46\arcsec$ (1.85 pc), larger than the largest angular size scale of these data, suggesting that there could be extended emission connecting the two northern clumps that has been resolved out.  This linear structure is apparent from 24.1 to 27.5 GHz, but it is absent at 36.4 GHz, likely because it falls in the less-sensitive outer regions of the primary beam at that frequency. As with the tangential filament  C9, the linear structure of C2 is also present at 3 mm \citep{Rathborne14b}. The morphology of this feature also appears to match that of an adjacent molecular gas filament seen in the \amm$\, maps presented in Section \ref{morph2}. We discuss this relationship further in Section \ref{cmorph}. 

In addition to the extended continuum emission discussed above, there are several smaller regions of more compact continuum emission located inside of the confines of the cloud, as traced by the molecular gas emission. These continuum regions (C1, C3, C4, and C6) are relatively bright and also have counterparts at higher frequencies: all are detected in the 90 GHz ALMA image of \cite{Rathborne14b}. The remaining compact regions, C5, and 10, are the the brightest sources detected at 24.1 GHz, and lie outside of the boundaries of the cloud traced by the molecular gas emission. These sources are only marginally resolved by the VLA observations, and their peak intensities at 24 GHz are greater than 1.0 m\jyb$, and are larger than the peak intensities of other sources in the cloud (except C1) by at least a factor of 2. Although they lie near the edge of the  \cite{Rathborne14b} ALMA map, they are still detected at 90 GHz. 

 A number of the compact ($<2\arcsec$) sources were also identified by \cite{Rodr13} using higher resolution VLA data. Their sources JVLA 1,4, and 6 correspond to the previously discussed sources C5, C1 and C10, respectively. In addition, they detect a source (JVLA 5), that is part of our more extended source C7. As \citet{Rodr13} also do not detect our C3, one of the stronger compact sources we detect in \thebrick, it seems likely that these higher resolution data are more strongly affected by spatial filtering (the largest angular scale recoverable in the B-configuration VLA data of \citet{Rodr13} should be around $10\arcsec$ ). It may then be that several of the apparently compact sources they detect are simply the peaks of intrinsically more extended structures. \cite{Rodr13} suggested these sources may represent high mass star formation associated with this cloud, in excess of that previously inferred by the presence of a single water maser in the northern part of the cloud. However, we note that in addition to several of their sources being associated with extended emission, all of them also lie outside of the dense molecular gas in the cloud, which is not what is expected if these sources are (proto)stellar in nature. We will discuss which of the compact sources we identify could be most consistent with star formation, based on their spectral indices, fluxes, morphologies, and relation to the molecular gas in the cloud, in Section \ref{cmorph}.

\subsection{Spectral Indices}
To determine the nature of the radio continuum sources, and particularly to judge the amount of ongoing star formation in \thebrick, the emission mechanism of the observed radio continuum in this cloud must be identified. We have calculated spectral indices for the ten representative regions identified above. However, the VLA data alone are found to be insufficient for calculating accurate spectral indices, as the small range of radio frequencies probed by these observations does not provide a large lever arm for determining spectral indices, and the combination of the extended nature of much of the emission and the location of many sources near the edge of the primary beam appear to make the fluxes derived at Ka band systematically low. For this reason, spectral indices are calculated using the fluxes at both K-band frequencies, and the 90 GHz fluxes from the ALMA continuum images of \cite{Rathborne14b}.  Separate spectral indices are calculated using the both the images made from just the ALMA 12m array, and the image that has been additionally corrected for missing extended flux via combination with single dish data. We expect that the true spectral indices likely fall between these two values, as our K band data are sensitive to larger angular scales than the ALMA data, and should thus recover more flux than the ALMA 12m only image, but less than the single-dish corrected ALMA image.

We find that the three sources at the center of the cloud (C3, C4, and C6) have consistently rising spectral indices, with $\alpha$ between 0.5 and 1.4, depending on whether the uncorrected or corrected ALMA fluxes are used in this calculation. The two sources outside of the cloud (C5 and C10) have consistently negative spectral indices: the spectral index of C5 is between -0.9 and -1.3, while the spectral index of C10 is between -0.3 and -0.7. This suggests that the emission from these two sources is dominated by nonthermal processes. This is in good agreement with \cite{Rodr13}, who find that the spectral indices between 1.3 and 5.6 cm for C5 and C10 (their sources JVLA 1 and JVLA 6, respectively) are $-0.9\pm$0.1 and $-0.3\pm$0.2, respectively. 

The remaining sources in \thebrick~(C1, C2, C7, C8 and C9), the majority of which are extended, have nearly flat spectral indices (ranging from slightly negative to slightly positive in the uncorrected and corrected ALMA images, respectively), consistent with free-free emission. It has been suggested that this cloud could be a good target for detecting limb-brightened synchrotron emission from cosmic ray interactions, as it previously showed few signs of free-free continuum emission that would confuse the synchrotron signal \citep{Jones14}. However, our observations show that there is actually significant free-free emission in \thebrick, and we do not see any indication of extended synchrotron emission. While we do detect continuum emission from the limb of \thebrick, it appears thermal in nature, and is only present on the eastern edge of the cloud. While lower frequency observations might prove more optimal for searching for extended synchrotron emission from this cloud, \thebrick~overlaps with a supernova remnant identified at 90 cm \citep{KF96}, and so ultimately this cloud is likely not an ideal candidate for detecting a synchrotron signal from cosmic ray interactions.

\subsection{Thermal Emission from Ionizing Photons}
\label{ion}

For these sources and the sources with rising spectral indices we then calculate the Lyman-continuum photon rate to further constrain the properties of this thermal emission and its origin. Assuming the continuum emission in \thebrick~is thermal in nature, caused by ionization from an external or embedded source, calculating the Lyman-continuum photon rate can give insight into the types of sources required to stimulate this emission. The number of ionizing Lyman-continuum photons needed to produce this emission can be calculated using the formulation of Mezger \& Henderson (1967): 

         \begin{equation} 
         N_{Lyc}=1.301\times10^{49}\left(\frac{T_{e}}{\textrm{K}}\right)^{-0.3}\left(\frac{S_{\nu}}{\textrm{Jy}}\right)\left(\frac{\textrm{D}}{\textrm{kpc}}\right)^{2}
	\label{lyc}
	\end{equation}
	\[   \times\left(\ln\left(\frac{0.0499 \textrm{GHz}}{\nu}\right)+1.5 \ln\left(\frac{T_{e}}{\textrm{K}}\right)\right)^{-1} \]

\begin{figure*}
\includegraphics[scale=0.5]{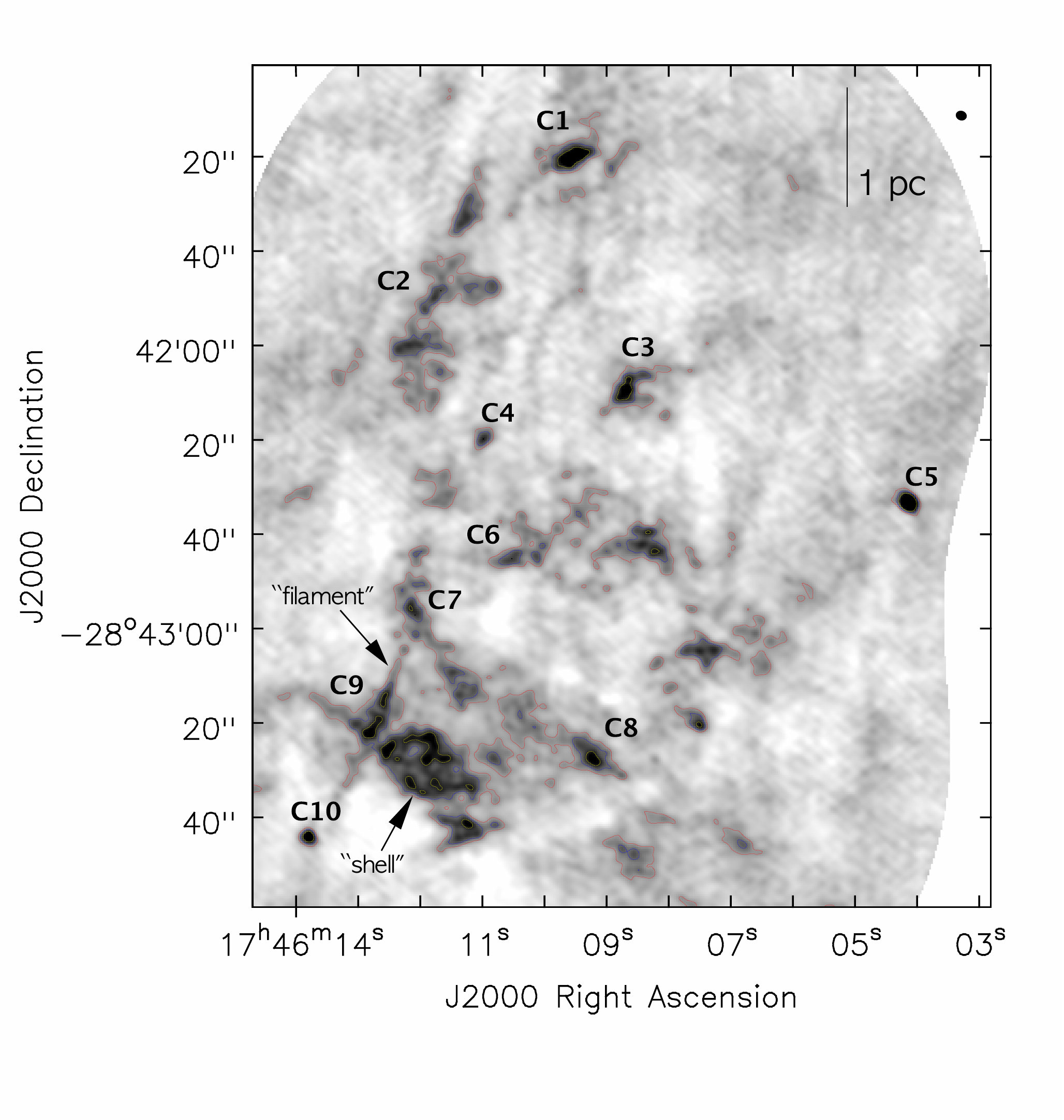}
\caption{1 GHz bandwidth continuum image at 24.1 GHz (top left panel in Figure 1) with 6, 10, and 15 $\sigma$ contour levels (from a non primary beam corrected rms of $3\times 10^{-5}$ \jyb$) colorized as red, blue, and yellow, respectively. The continuum regions of interest are labeled as C1-10, using the specified contour levels stated in Table 3, depending on the compactness of the emission source. The region parameters, Lyman continuum flux, and spectral indices for these regions are also presented in Table 3. }
\label{findingchart}
\end{figure*}

Where T$_{e}$ is the electron temperature, assumed to be 10,000 K, S$_{\nu}$ is the flux density at a frequency $\nu$ in GHz, and D is the distance to the Galactic center, assumed to be 8.4 kpc. The Lyman-continuum photon rate was calculated for all 10 of the sources in Figure \ref{findingchart} (except for C5 and C10, which have a nonthermal spectral index at these frequencies) using the flux densities at 24.1 GHz, and are presented in column 12 of Table \ref{SItab}. The tabulated values of Log N$_{L}$ range from 45.9 to 47.5 with the largest values ($>$ 46.7) corresponding to the large diffuse regions. Assuming the stars producing this ionization are on the zero-age main sequence, which would be expected for a cloud undergoing star formation, this range of Log N$_{L}$ values would correspond to ionization by a single star with a spectral type of B1 to O9.5 (Panagia 1973). The three compact regions (C3, C4, and C6) located towards the center of the cloud would each be ionized with a star of spectral type B0.5. However, as these latter sources have apparently rising spectral indices, it is likely that the inferred Lyman continuum fluxes are either systematically overestimated (if these fluxes are contaminated with dust emission) or systematically underestimated (if these sources are in fact optically thick). As we will discuss further in Section \ref{cmorph}, it is not possible to determine which is more likely. 

 \section{Molecular Gas Morphology and Kinematics}    
 \label{morph}
Not only does the large bandwidth of these VLA observations make possible the first sensitive radio continuum map of \thebrick, it also enables a survey for spectral line emission in the cloud over a total bandwidth of $\sim$4 GHz. In total, we detect and image 12 molecular lines from 4 species in \thebrick. With the high spectral (1-3 \kms$) and spatial (3$\arcsec$ = 0.1 pc) resolution of these observations, it is possible to investigate the detailed morphology and kinematics of the molecular gas in \thebrick, and for the first time to compare the distribution of the continuum emission from ionized and nonthermal structures with that of the molecular gas. 

\subsection{Morphology}         
\label{morph2}

\begin{figure*}
\includegraphics[scale=1.0]{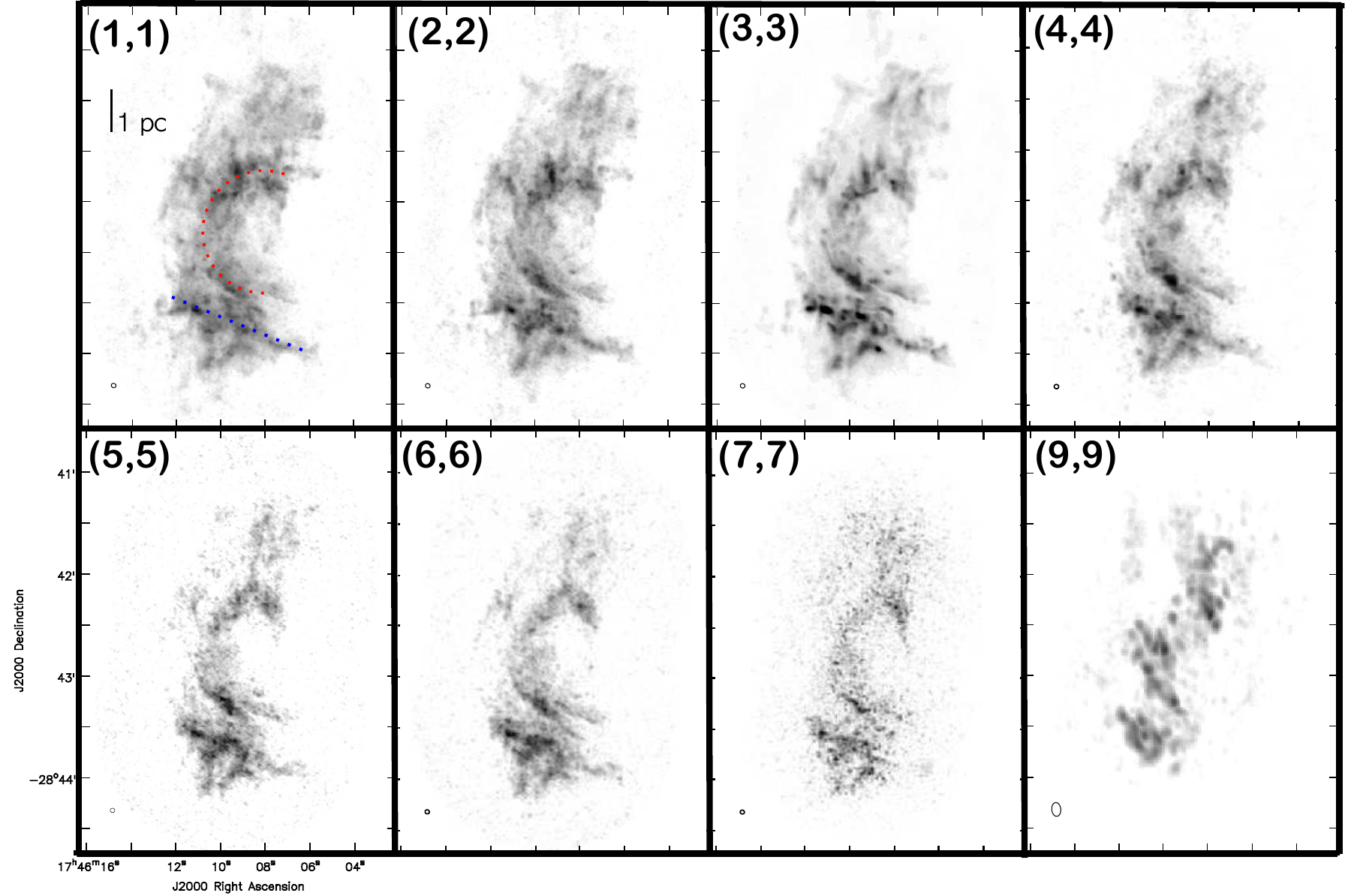}
\caption{ Maps of the maximum intensity over all velocities for the \am~(1,1) to (9,9) transitions. The morphological features referred to as the \carc and the ``tilted bar'' are indicated in the (1,1) map as a dotted red curve and a dotted blue line, respectively. }
\label{M8}
\end{figure*}

The majority of the observed lines in \thebrick~(8/12) are from ammonia (\am\,). Figure \ref{M8} shows the peak intensity of these 8 observed transitions of \am: ($J,K$) =  (1,1), (2,2), (3,3), (4,4), (5,5), (6,6), (7,7), and (9,9) (see Table \ref{Imaging} for image parameters). The first 7 lines were imaged simultaneously with a single correlator setting at K band, while the (9,9) was observed separately at Ka band. As the emission from the (9,9) line is much weaker than the others, it was smoothed to improve the imaging signal to noise. The strongest observed line is the (3,3) line (typically about twice as bright as the (1,1) or (2,2) lines). We take it to be generally representative of the distribution of \am\, in this cloud given that, as can be seen in Figure \ref{M8}, all of the observed \am\, transitions exhibit very similar structure. In general, the emission is diffuse and filamentary throughout the cloud with many curved features. In the strongest lines (J,K $\leq4$), much of the emission can be seen to be concentrated in a number of compact clumps. These clumps are most prominent in the (3,3) line, likely in part because this line is the strongest. The clumps have typical brightness temperatures of 10-60 K in the (3,3) line which is consistent with thermal emission, although (3,3) masers have been previously suggested to exist in CMZ clouds \citep{JMP99}. 

In our \am\, maps, we identify two primary features of the morphology of \thebrick. emission. The first is an apparently ``C'' shaped arc located near the center of the cloud (hereafter ``C-arc"). The \carc~extends roughly $90\arcsec$ (or 3.7 pc) in declination (from $-28\degr 42\arcmin 00\arcsec$ to $-28\degr 43\arcmin 30\arcsec$, at an RA of $\sim$17\h46\m08\s$).   The \carc~structure can be seen in recent millimeter spectral line studies of this cloud using ALMA \citep{Higuchi14,Rathborne15}, and the SMA \citep{Johnston14}. It is suggested by \cite{Higuchi14} that this feature is the remnant of a recent collision between \thebrick~and a smaller cloud. The \carc\, has roughly the same brightness as emission in other regions of the cloud in the (1,1) and (2,2) lines (likely because the bulk of the emission in these lines is optically thick, as will be discussed further in a subsequent paper) but it is prominent in the higher-excitation lines of \am. Intriguingly, this arc follows very well the direction of the magnetic vectors inferred from recent polarization studies of this cloud \citep{Pillai15}.

The second feature is a ``tilted bar" below the \carc, beginning on the eastern side of the cloud at a declination of $-28\degr43\arcmin.5$ and spanning nearly the entire width of the cloud in right ascension. In lines of \am\, (3,3) and above, this region and the adjacent southern portion of the C-arc are the sources of the most intense \am\, emission in \thebrick, and this tilted bar contains the bulk of the brightest clumps seen in the \am\, (3,3) line. In addition to the prominent tilted bar in the southern half of the cloud, there are also a number of weaker linear emission features in both the north and south of the cloud. The easiest to identify in the \am\, maximum emission maps is a feature on the eastern edge of the cloud above the \carc. It is narrow, extending from a declination of $41.25'$ to $41.75'$, and is best seen in the K $\leq$ 4 lines. The southern edge of the \carc\ and the southwestern ``tail" of the bar also form nearly linear elongations. We will return to a discussion of the linear emission features in \thebrick~and how they relate to the linear radio continuum features and the HCO$^+$ absorption filaments identified by \cite{Bally14} in Section \ref{cmorph}.

\begin{figure*}
\includegraphics[scale=0.95]{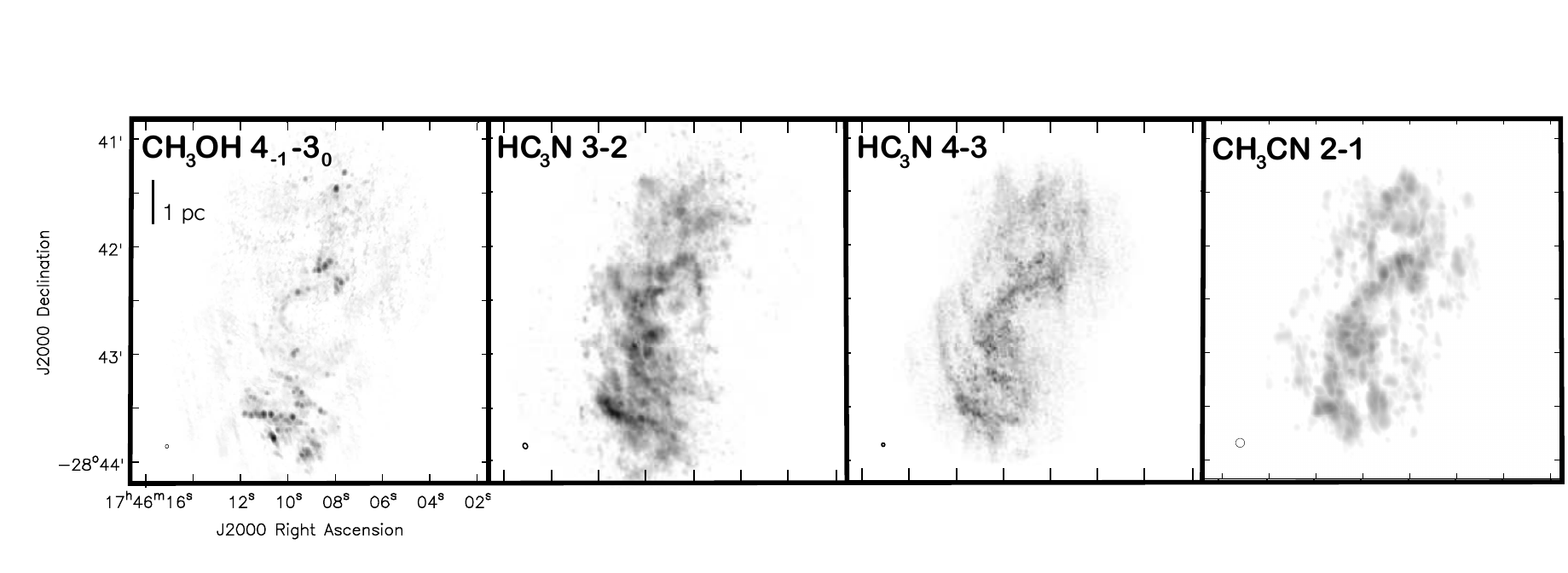}
\caption{Maps of the maximum intensity over all velocities for the {\bf Left:} \meth\, $4_{-1}-3_{0}$ line {\bf Middle-Left:}  \cyano\, 3-2 line  {\bf Middle-Right:} \cyano\, 4-3 line and  {\bf Right:} the blended CH$_3$CN 2$_k$-1$_k$ doublet in \thebrick.}
\label{Cyano}
\end{figure*}

In addition to the \am\, lines, at Ka-band, we also detect the ($4_{-1}-3_{0}$) line of \meth\, and the  (3$-$2) and (4$-$3) transitions of \cyano~and the (2$-1$) doublet of \methcy, maximum emission maps of which are shown in Figure \ref{Cyano}. The \meth\, line shows the most striking difference in morphology compared to all other lines imaged in this cloud: it is primarily composed of emission from dozens of discrete point sources. More than half of these point sources are located in the southern bar. The majority of the observed point sources have brightness temperatures $>$400 K and are likely masers;  we discuss the nature of the observed 36.2 GHz \meth\, emission sources in \thebrick~further in Section \ref{mas}.  Overall, these molecules trace the same general structure as seen in \am: the \carc is clearly visible, and the southern bar can be seen in the \meth\,  and \cyano\, lines, though it is less prominent in the faint line \methcy\, line. However, the \cyano\, and \methcy\, images both exhibit stronger emission at the center of the cloud (along the \carc, between a declination of $-28\degr42\arcmin.5$ and $-28\degr43\arcmin$) than is seen in either the \am\, or \meth\, images.  As can be seen in millimeter continuum images \citep{Rathborne14b,Johnston14}, the center of the cloud is also where the dust continuum emission is strongest, suggesting that it is the location of the densest gas. 

\subsection{Kinematics}         
\label{kin}

\begin{figure*}
\includegraphics[scale=0.6]{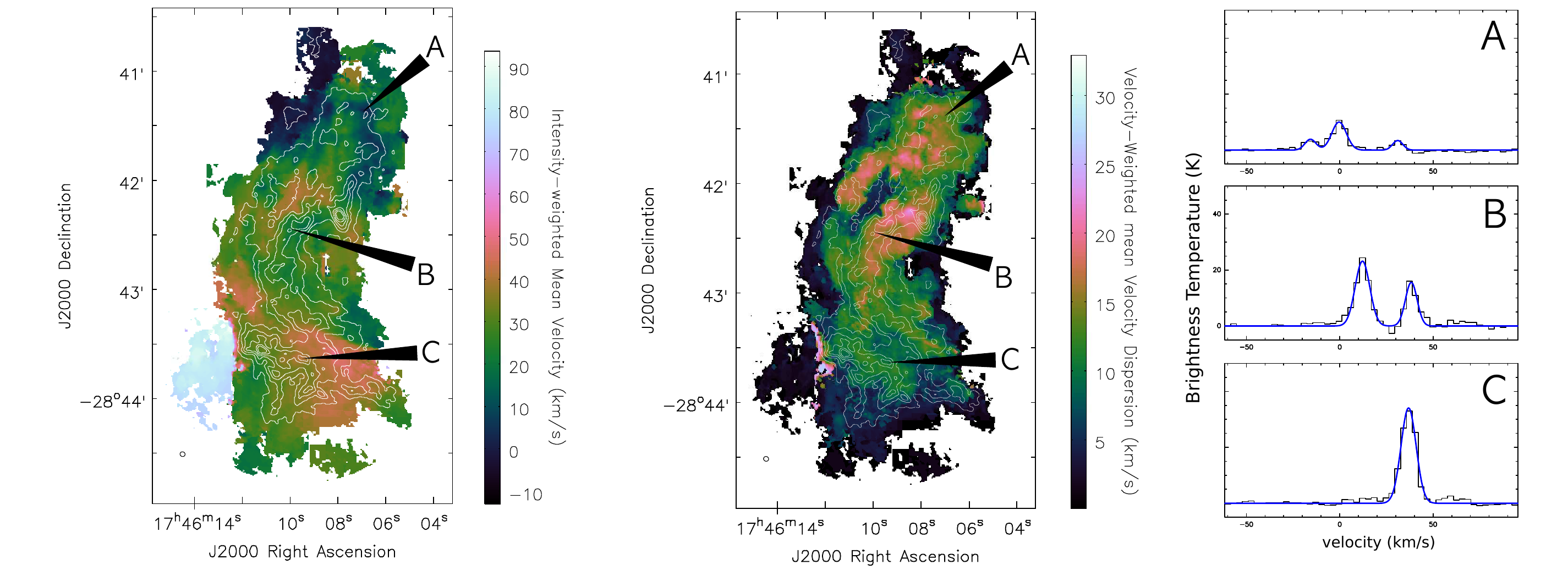}
\caption{ {\bf Left:} A map of the intensity-weighted velocity (moment 1) from the \am~(3,3) line. {\bf Center:} A map of the intensity-weighted velocity dispersion (moment 2) from the \am~(3,3) line. Overlaid on both the moment 1 and 2 images are contours of the integrated value of the spectrum (moment 0). {\bf Right:} Spectra extracted from the \am\, (3,3) cube from 3 positions in \thebrick}
\label{velspec}
\end{figure*}

Although \thebrick~may be considered quiescent in terms of its (lack of) ongoing star formation activity, its kinematics are much more active. In the left panel of Figure \ref{velspec}, we show a Moment 1 map of the intensity-weighted velocity from the \am\, (3,3) line, which is the brightest of the observed \am\, transitions. The map was made by limiting the emission spatially to the region previously identified during the initial CLEAN, and by using a threshold of 2$-$3 times the rms noise value for the spectral line. Emission toward the cloud in the (3,3) line spans velocities from -10 \kms$ to 90 \kms$ (with weak emission, which does not contribute significantly to the average values in this figure, extending to velocities as low as -40 \kms$). The lowest velocities (-10 to +20 \kms$) in the cloud generally fall in the northern region of the cloud, while the southern region of the cloud is characterized by gas in the range of 20 to 60 \kms$. The highest velocities (70$-$90 \kms$) are primarily confined to a region at the southwest edge of \thebrick, which has been suggested by \cite{Rathborne14a} and \cite{Johnston14} to be a separate cloud which may or may not be related to the main cloud. 

A similarly-constructed map of the intensity-weighted velocity dispersion (Moment 2) in the same line (Figure \ref{velspec}, middle panel) shows dispersions ranging from 2 to 30 \kms$. However, the largest of these velocity dispersions ($\Delta$v $>$ 20 \kms$) may be misleading due to the presence of multiple components along some lines of sight. This complicated velocity structure of \thebrick~is represented in three example spectra shown in the rightmost panel of Figure \ref{velspec}. There are multiple velocity components in the northern part of the cloud along the same line of sight that confuse this analysis, causing the velocity dispersion to represent a combination of the width of these components and their separation. An example of this is shown in spectrum A, a triple profile spectrum with intensity peaks around -15, 0, and 30 \kms$. Areas with multiple velocity components can be similarly poorly represented in the Moment 1 map: often the average velocity lies between velocity peaks, and is located at a velocity which little or no gas is present. Spectra towards the center of \bricks, along the \carc, discussed in \S 4.1, show a double peak profile as can be seen in spectrum B of Figure \ref{velspec}. The peak intensities of the profiles presented in spectrum B fall at velocities of 10-15 and 35-40 \kms$. This kinematic structure is typical for regions along the \carc. The southern part of \thebrick~has single line profiles (see spectrum C), with velocities greater than 30 \kms$, that typically represent the brightest emission in the cloud, with brightness temperatures at least 2-3 times the emission from other parts of the cloud. 

\section{Widespread 36.2 GHz \meth\, Maser Emission}
\label{mas}

As discussed in Section \ref{morph2}, the 36.2 GHz \meth~($4_{-1}-3_{0}$) line exhibits the most unique morphology of all of the spectral lines we observe. While all of the other lines exhibit extended, filamentary, and clumpy structure, the \meth\, emission lacks extended structure, and consists nearly entirely of discrete point sources. We detect dozens of these point sources, shown in Figure \ref{masermax}, a maximum intensity map of the \meth\, emission within the cloud.  While the distribution of sources spans the entire cloud, the majority of the sources, including nearly all of the brightest sources, are concentrated in the southern part of \thebrick. 

The point-like nature of the bulk of the \meth\, emission is consistent with the $4_{-1}-3_{0}$ transition being a well-documented ``class I" or collisionally-excited \meth\, maser \citep{Morimoto85,Menten91a,Slysh94,Sj10}. In general, masers in this line are observed to trace shocks: they are found in outflows in early stages of both low and high-mass star formation \citep{Kurtz04,Chen09,Kalenskii10}, around expanding ultracompact HII regions \citep{Voronkov10}, as well as in the shells of expanding supernova remnants interacting with molecular clouds \citep{Pihlstrom14}.  Although within the plane of our Galaxy, class I \meth~ masers have thus far been observed to be associated nearly exclusively with early stages of star formation, in the CMZ this association is less clear. In Sgr B2, roughly a dozen 44 GHz  ($7_0-6_1$) class I masers are observed, many of which are not near known sites of star formation in the cloud \citep{MM97}. These masers have been suggested to be induced by large-scale shocks from a cloud-cloud collision, which has also been suggested to excite 36 GHz masers observed near Sgr A \citep{Sj10}. \cite{MM97} also observe quasithermal emission in the 44 GHz line in Sgr B2, which is interpreted as originating in denser gas in which the maser has been quenched \citep{Menten91a}.   More recently, \cite{YZ13} find emission from the 36 GHz line to be widespread in the CMZ. The large number ($>$ 350) of these sources is perhaps not surprising, as models \citep[e.g.,][]{Cragg92} suggest that the 36 and 44 GHz \meth\, transitions are the easiest of 28 known and predicted class I \meth\, masers to excite \citep[though new class I \meth\, masers continue to be predicted and detected;][]{Voronkov12, Yanagida14}. The widespread distribution of these sources in the absence of other tracers of widespread star formation leads \cite{YZ13} to dismiss this as the likely origin of the \meth\, emission (nor is it likely to be due to supernova interactions, as the sources are not confined to the few known supernova remnants in the region). Instead, \cite{YZ13} suggest that the enhanced \meth\, abundances in in the CMZ are a result of desorption from grains by cosmic rays. We discuss the merits of both a shock and cosmic ray model for giving rise to these \meth\, sources in Section \ref{methshock}. Ultimately however, the low spectral resolution ($\sim17$ \kms$) of the \cite{YZ13} survey does not allow for the 36 GHz sources to be positively confirmed to be masers. 

\subsection{Identification of Maser Candidates and Source Catalog}

Although we observe the \meth\, line with similar spatial resolution as \cite{YZ13}, our spectral resolution of  $\sim1$ \kms$ is better able to discern whether these sources have nonthermal brightness temperatures. In order to more quantitatively analyze the observed \meth\, emission in \thebrick~and determine whether the observed sources are masers, we have produced an initial catalog listing the properties of the strongest detected sources. 

The \meth\, emission seen in Figure \ref{masermax} is clustered together both spatially and spectrally.  This clustering makes manually distinguishing between individual sources difficult. In order to examine these complicated fields, we adopt a version of the source detection algorithm {\it Clumpfind} \citep{Williams94} which distinguishes between sources that may partially overlap in position or velocity. {\it Clumpfind} identifies local maxima, then examines the emission surrounding the maxima both spatially and spectrally to determine the boundaries of the source. No assumptions about the clump geometry, neither spatially or spectrally, are made during processing by the algorithm. {\it Clumpfind} produces a list of maser candidate clumps with uniform criteria. The output of {\it Clumpfind} is  then used to construct the catalog. Since a significant portion of the maser emission lies near the edge of the observed Ka-band field, a primary beam correction was applied while calculating the properties of the sources found by {\it Clumpfind}. Clumpfind searched for emission down to six times the RMS noise in each channel. 

Our {\it Clumpfind} analysis of \thebrick~yields 383 \meth\, clumps with a brightness above six times the RMS noise. However, in order to remove the possibility of false detections, we required sources to have a brightness greater then ten times the RMS noise in their spectral channel in order to be included in the catalog; 195 \meth\, clumps meet this criterion, which is a conservative cut-off that ensures that we are examining masers and not artifacts from several of the extremely bright masers in the field. However, as a result the final catalog of sources is incomplete below a flux of 1.0 Jy (the largest residual in the cube after removing the {\it Clumpfind}-detected sources), with the incompleteness being most significant near the velocity range of the brightest masers. Additionally, we make a total flux cut at 0.3 Jy, below which emission structure in the image begins to become significantly compromised by the missing flux on large scales. This removes 47 more sources, leaving a total of 148 detected \meth\, point sources. 

These {\it Clumpfind} sources are then divided into two catalogs: masers, and candidate masers. Of the 148 catalogued point sources, 68 have brightness temperatures $>$ 400 K, in excess of the highest gas temperatures suggested to exist in this cloud \citep[$\sim$ 325 K,][]{Mills13,Johnston14}. This indicates that they are likely nonthermal, and we classify them as masers. Their properties, including the FWHM line width and peak brightness temperature, are given in Table \ref{Masers}.  The remaining 80 sources have brightness temperatures that could be thermal, and so cannot yet be confirmed to be masers, given the limited spatial and spectral resolution of these data. Their properties are given in Table \ref{Candidates}.  Spectra for all of the catalogued \meth\, sources, both masers and candidates, are presented in Figures \ref{maserspect} and \ref{candspect1}. Due to the spatial and spectral clustering of the sources, the spectrum of a candidate may show other peaks from bright masers located nearby.  To aid in identification of weak masers near brighter sources, the central velocity of each maser candidate is indicated by a dashed line in each spectrum.

More than half of both the masers (37; 54\%) and maser candidates (43; 54\%) are spatially unresolved. These sources are deemed to be spatially resolved if their FWHMs are larger than twice the synthesized beam area. All of the masers and maser candidates also have relatively narrow line widths, with a mean FWHM for all catalogued sources of 3.6 \kms$. More of the candidates (28; 35\%) are spectrally unresolved than the masers (10;15\%), with sources deemed to be spectrally resolved if their FWHMs are larger than twice the channel resolution of 1.02 \kms$. As we expect masers to have subthermal ($<$1 \kms$) linewidths and generally to be spatially unresolved point sources, the large fraction of masers that are both spectrally and spatially resolved suggests that there is still confusion in this catalog, and that our observations are still underestimating the true number of maser sources in this cloud. Higher-resolution VLA observations should be able to confirm this and to determine the clumping properties of the \meth\, masers in this cloud. 

As the maser candidates appear, apart from their lower brightness temperatures, to be quantitatively similar to the masers (with similar fractions of both sources spatially and spectrally unresolved) we expect that the bulk of these sources will also prove to be masers. Possible exceptions to this are some of the candidate sources that have broader lines and more spatially extended emission. These sources (as well as regions of extended emission with T$_B<< 100$ K that are not included in our catalogs) could represent thermal or ''quasithermal'' emission, as we discuss further below. 
 
The number of \meth\, sources detected in \thebrick~in our observations (148) is thus far unprecedented for a CMZ cloud. It is larger than the number of 36 GHz masers (10) recently identified by \cite{Sj10} in the 50 and 20 \kms$ CMZ clouds, however this difference in the number of sources may be due in part to the fact that our observations are $\sim5\times$ times more sensitive. It is also more than an order of magnitude greater than the number of 36 GHz sources (8) previously identified in this cloud by \citet{YZ13}. However, we should note that the positions for these sources given by \citet{YZ13} do not match the positions of any of our masers, and it appears upon checking the archival data for these observations, that the previously published positions of 36 GHz sources in this cloud are incorrect. In the central part of the cloud, where we compare our data to the archival data from the \citet{YZ13} observations, we detect 3 sources, at the positions of our brightest masers (M10, M18 and M25), but nothing at the published positions of three sources in this field (catalog numbers 42,43,44).  We will assume that the number of sources catalogued by \citet{YZ13} in this cloud can still be taken to be order-of-magnitude representative of what can be detected in this cloud at the sensitivity of their survey. Based on the yield of the \citet{YZ13} study-- 356 individual 36 GHz sources over a surveyed area of 0.33 square degrees-- our observations then suggest that clouds in the CMZ could host thousands of these masers, detectable by observing more clouds in the same way as for the \thebrick~data presented here, or potentially with a higher spectral-resolution survey than that conducted by \cite{YZ13}.

\subsection{Distribution of masers in \thebrick}
\label{methcompare}

\begin{figure*}
\includegraphics[scale=0.52]{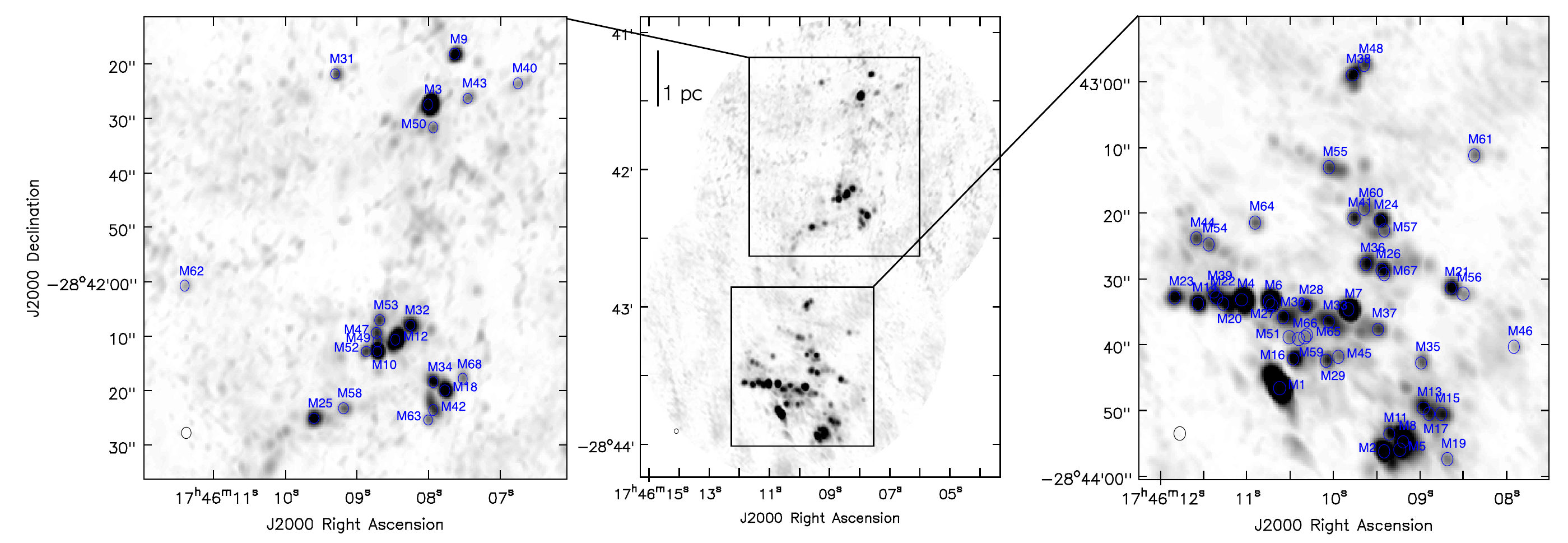}\\
\includegraphics[scale=0.44]{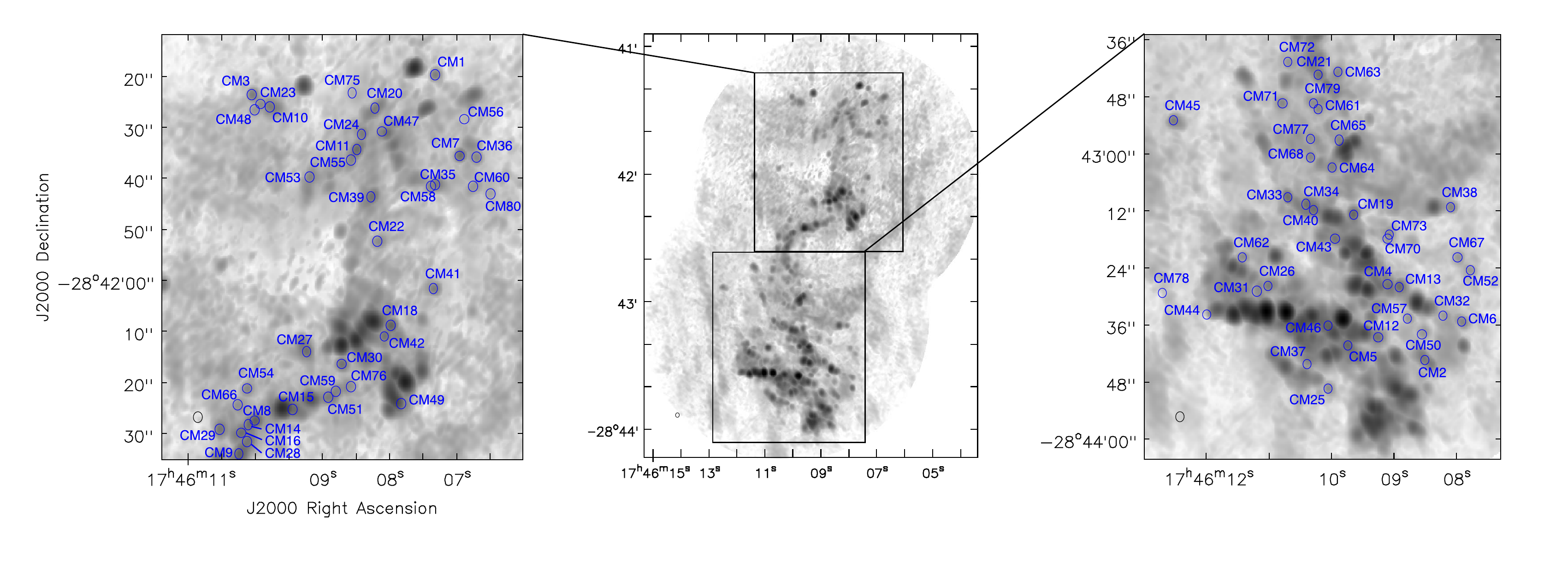}
\caption{ Maps of the peak 36.2 GHz  \meth\, emission detected in \thebrick. {\bf Top:} A map of the \meth\, sources which we have determined to be masers. Sources are labelled M1-M68, and their properties are given in Table \ref{Masers}. {\bf Bottom:} A map of the \meth\, sources which cannot yet be determined to be masers, and are catalogued as maser candidates. In order to make visible the emission from these weaker sources, noisier channels in the cube which contain the two brightest masers (corresponding to velocities of 28-32 \kms$) were not used to construct this image. The candidate sources are labelled CM1-CM80, and their properties are reported in Table \ref{Candidates}. }
\label{masermax}
\end{figure*}

\begin{figure}
\includegraphics[scale=0.85]{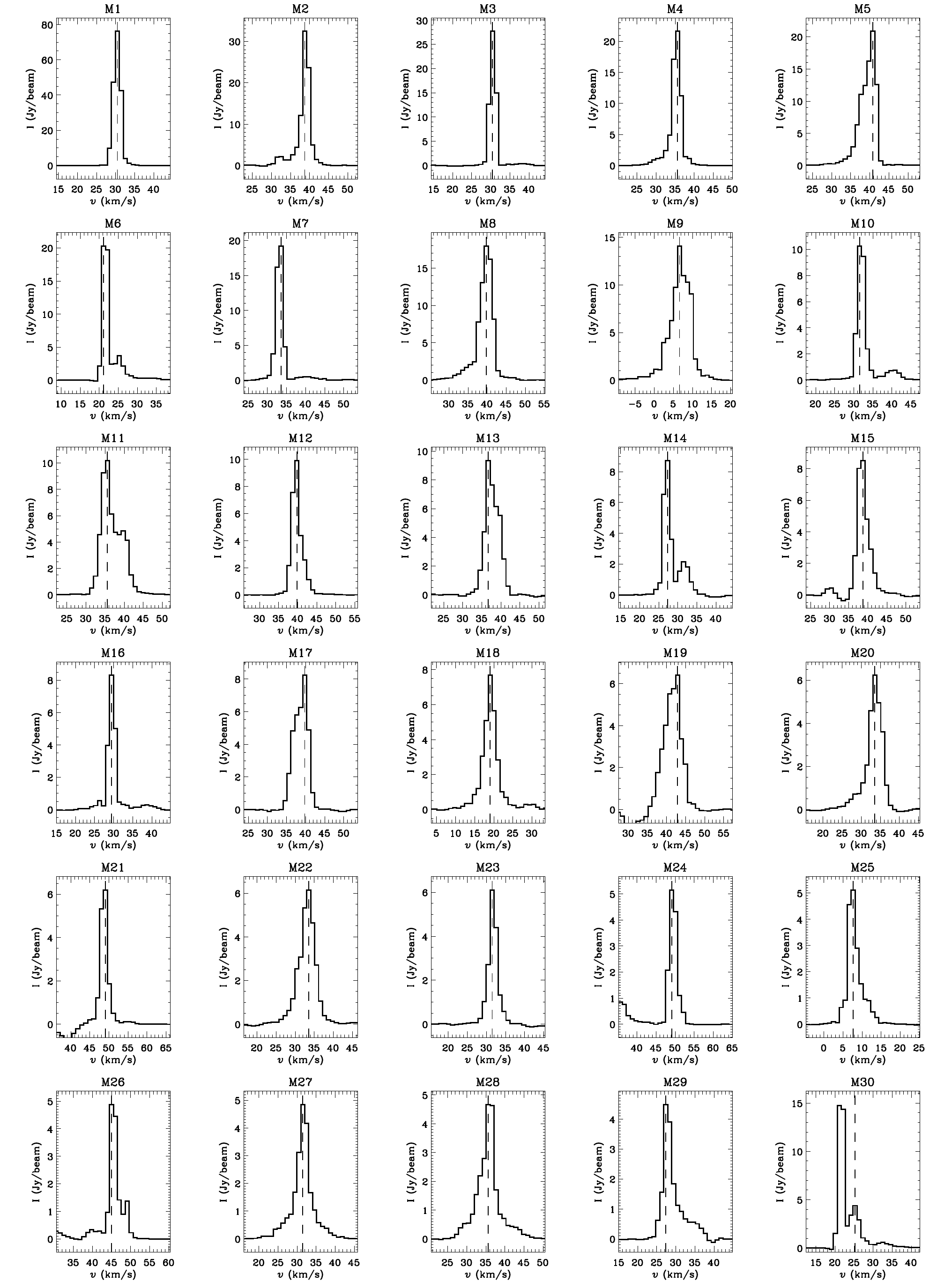}
\caption{ Spectra of all catalogued masers. Each spectrum corresponds to the pixel associated with the peak emission in the source defined by {\it Clumpfind}. The conversion factor to go from Jy to K is 250.5}
\label{maserspect}
\end{figure}

Focusing on the subset of sources that we can confirm to be masers, it can be seen from Figure \ref{masermax} that these are distributed throughout the entire cloud. However, 44 of the masers (more then 60\% of the total) are concentrated in the southern regions of the cloud, south of declination$ = -28$\degr43\arcmin00.0\arcsec (See Figure \ref{masermax} Left). These regions correspond to the regions denoted from the ammonia maps as the \carc and the ``bar''(see Section \ref{morph}).  Masers in the northern part of the cloud are typically more isolated than masers that fall in these other two regions. In addition to containing the majority of the maser emission, the southern region of the cloud also contains the brightest maser emission.  With the exception of M3, all of the brightest masers are found in the southern region of the cloud.  These brightest sources (M1 through M8) exhibit brightness temperatures in excess of 4000 K. In addition to the seven brightest masers, 25 additional masers have brightness temperatures in excess of 1000 K.  The velocity range of the \meth~masers is from $-5$ \kms$ to 50 \kms$.  Masers with velocities less than 20 \kms$ are seen only in the northern portion of the cloud, while masers at velocities greater than 20 \kms$ are seen throughout the cloud. 

In general, the velocity distribution of the masers ($-5$ \kms$ to 50 \kms$) follows the same velocity distribution traced by \am~(3,3) and seen in Figure \ref{velspec}. The northern part of the cloud shows maser emission occurring at roughly two velocities, $\sim$10 \kms$ and 35 \kms$, while the southern part of the cloud has a single, higher velocity component, $\sim$35 \kms$. These maser velocities are similar to the \am~(3,3) velocities, discussed in Section \ref{kin}. The brightest masers, typically found in the southern part of the cloud, have velocities between 30 \kms$ and 40 \kms$, which is also the velocity of the brightest \am~(3,3) emission. 

The majority of the \meth~ masers do not correspond to any continuum features, though a few exceptions are seen. The rising-spectrum continuum source C3 and the shell-like continuum source C9 are both associated with maser emission.  The masers located in the vicinity of C3 ( M10, M12, M32, M47, M49, M52, M53) are clustered around the continuum source, near a velocity of 40 \kms$. We further discuss possibilities for the nature of this region, which is also associated with a peak in the millimeter dust continuum, in Section \ref{dis}. The masers located in the vicinity of the shell source (M4, M20, M22, M14, M23), together with M6, M7, M27, M28, M30 and M33, form a nearly straight line at roughly constant declination across the cloud. This linear feature corresponds to the northeastern edge of the ``bar'' feature seen in \am\, and discussed in Section \ref{morph}. Two weaker masers (M44 and M54) are located near the top of the western edge of the shell.

The spatial distribution of the \meth~masers appears extremely similar to that of the dense gas traced by \am\, (especially the \am\, (3,3) line) in \thebrick. Like the morphology of the \am\, lines in Figure \ref{M8} which are brightest in the southern part of the cloud, the majority of the brightest \meth~masers are also observed to be in the southern part of the cloud, and are associated with several bright, compact regions of \am. The \meth~masers also appear to be coincident with other prominent features traced in \am~(3,3), such as the \carc~and, as previously mentioned, the ``bar''. A close correlation between \meth~and \am~emission, especially in the (3,3) line of \am\,, has been previously noted for gas clouds in the GC \citep[e.g., M$-$0.02$-$0.07,][]{Sj10}. In other star forming regions in the Galaxy, masers in the (3,3) line of \am\, have been observed to arise in the same region as collisionally-excited \meth\, masers \citep[e.g.,]{Mangum94}. While \am\, (3,3) masers have been suggested to exist in the CMZ cloud Sgr B2 \citep{JMP99}, in \thebrick~all of the (3,3) emission has brightness temperatures $<$100 K and so cannot be clearly attributed to masers. Finally, no \meth\, emission is observed at the location of the H$_2$O maser identified by \cite{Lis94}. 

In general, the maser candidates follow the same distribution as the masers: distributed throughout the cloud, with the majority lying in the southern half. There are also two \meth~ maser candidates associated with the faint, 80 \kms$ component in the south-east portion of the cloud (CM44 and CM78). A number of the candidate sources (e.g., CM8, CM9, CM14, CM15, CM16, CM21 and CM28) also trace out a crescent-like feature the center of the cloud corresponding to the \am~ \carc. The typical brightness temperatures of these sources are 200-300 K, and they tend to have somewhat broader than normal measured FWHMs: $\sim$ 4.5-5 \kms$. In addition to this main peak, the properties of which are catalogued, many of these spectra also exhibit a weaker superposed component having brightness temperatures of 40-80 K, and linewidths of 6-10 \kms$, which appears as a plateau or `wings' in the spectra of many of these candidates (e.g, CM16). This weak and apparently spatially extended \meth\, emission appears to be primarily associated with an analogous `c'-shaped feature in the 3 mm ALMA dust continuum map of \cite{Rathborne14b}. The low brightness temperature of the extended emission and its close correspondence with the 3 mm continuum could indicate that this region contains ``quasithermal'' emission from gas sufficiently dense that the maser in this line is quenched, analogous to that seen in the 44 GHz line in Sgr B2 \citep{MM97}. If emission in this region is quasithermal in nature, then it is likely to be seen as a maser in other \meth\, transitions which quench at higher densities \citep[e.g., 44, 84, or 96 GHz;][]{Cragg92,McEwen14}.

\section{Is there Ongoing Star Formation in \thebrick?}
\label{dis} 

\subsection{Origin of the \meth\, Masers in \thebrick}
\label{methshock}

In the interstellar medium, \meth\, and other ``saturated" (hydrogen-rich) molecules are primarily believed to be formed on the surface of dust grains \citep{TH82,Charnley92,WK02}, where \meth~and \am\, are among the most abundant mantle species present relative to H$_2$O, as measured in both low and high-mass YSOs and cold cloud cores \citep{TA87,Dartois99,PvDD04,Gibb04,Boogert08,Oberg11}. Notably, for \meth, the high abundances measured for maser sources are inconsistent with those predicted by gas-phase formation models \citep{Menten86,Hartquist95}. To get the \meth\, off of the dust grains and into the gas phase in the observed large quantities then requires a mechanism to liberate the \meth\, from the grain mantles. Proposed mechanisms include thermal desorption via heating from an embedded protostar, shocks, or cosmic rays \citep[requiring grain temperatures $\gtrsim$ 90 K][]{Tielens95,BB07}, or nonthermal desorption processes including photodesorption via far-UV photons from cosmic ray interactions \citep{Prasad83,DAG85,Oberg09b}, grain sputtering, wherein the ice mantles of grains are dislodged via collisions (often in shocks) with other grains, neutrals, ions, or cosmic rays \citep{Johnson91,Caselli97}, and finally exothermic chemical reactions on the grain surfaces \cite{DW93,Roberts07}. In particular, the presence of molecules formed on grains in relatively cool and dense environments requires an efficient nonthermal desorption process \citep{WW93,Roberts07,Oberg09a,Caselli12}. In the CMZ, the abundance of 36 GHz \meth\, sources has been suggested to be due to photodesorption from cosmic rays in this region \cite{YZ13}. We reconsider this in the light of our new, high-resolution observations. 

Our observations reveal that locations of \meth\, emission are in general an excellent match to the 3 mm ALMA dust continuum map shown in \cite{Rathborne14b}. However, the observed \meth\, masers are stronger and more numerous in the southern part of the cloud, while the stronger dust emission is found in the northern portion of the cloud, north of Declination -28:42:34.2.  If high column densities of \meth\, simply originate from high column densities of dust, and if all of the excitation conditions are uniform, one might expect more masers in the northern parts of the cloud \citep[we also note that no \meth\, emission-- thermal or maser-- is detected toward the two strongest millimeter continuum peaks identified in][]{Rathborne14b}. One possible explanation might be that, for much of the northern part of \thebrick, the \meth\, emission is quasithermal, and the masers are quenched in those regions which correspond to not just high column densities but high volume densities. With future observations, it should be possible to test this with observations of more highly-excited masers (e.g., 44 GHz) which are quenched at higher densities \citep{Cragg92,MM97,McEwen14}. 

Another possibility is that the differences in the distribution and strengths of the masers are a result of variations in the geometry and kinematics of the cloud. Maser emission requires a velocity-coherent path length of gas for amplification of the emission. Class I \meth\, masers are, for example, rarely seen in the high-velocity components of outflows, which is suggested to be because the longest gain paths are found perpendicular to the outflow, at velocities near the systemic values \citep[][although as noted by \citealt{Voronkov14} this is also partially a selection effect]{Menten91a}. This makes it somewhat surprising to see a large quantity of masers in an extremely turbulent environment like that of \thebrick. However, the vast majority of the detected masers are relatively weak (having intensities $<$ 5 Jy) which could be a result of the short coherent path lengths in this gas. The stronger masers observed in the southern parts of the cloud could be due to a geometrical effect, larger gain lengths can be had perpendicular to the motion of a shock front \citep{KN96}, so if these masers trace a shock propagating in the plane of the sky, it could explain their enhanced intensity. However, are shocks really the mechanism responsible for generating these masers? 

Prior observations would seem to be able to rule out thermal desorption processes for clouds in the CMZ like \thebrick~which lack advanced stages of star formation.  Measured dust temperatures in \thebrick~are $<$ 30 K \citep{Molinari11,Longmore12}. Dust heating via cosmic rays should be relatively uniform (though it may be more efficient toward the edges of the cloud), and is further not predicted by models to yield dust temperatures above $>$40 K in \thebrick~\citep{Clark13}, so thermal desorption via cosmic ray heating can be ruled out. Heating via shocks or embedded sources might lead to discrete regions of higher dust temperatures, however the dust temperature maps of \cite{Longmore12} show no signs of temperature variation that might indicate unresolved regions of higher temperatures. 
Although there is one location, C3,  where the observed \meth\, masers cluster around a bright radio continuum source, the masers generally do not correlate with the radio continuum, making the heating of dust from embedded sources an unlikely source for the liberated \meth~more globally observed in the widespread masers in this cloud. In general, the distribution and relatively low intensities of the continuum emission do not suggest that there are embedded, ionizing sources within this cloud (see below). 

Among the previously listed nonthermal desorption processes, those most likely to be important in the unique environment of dense clouds in the CMZ are then sputtering from shocks and UV photodesorption due to cosmic rays. There is evidence in the CMZ for enhanced rates of both of these processes: both a cosmic ray ionization rate several orders of magnitude greater than the local value in the solar neighborhood \citep[$\zeta_{GC} = 10^{-15}-10^{-13}$ s$^{-1}$;][but see also \citealt{vanderTak06} who find $\zeta_{GC} \sim 10^{-16}$ s$^{-1}$ in Sgr B2]{Dalgarno06,Goto13,YZ13b,YZ13c,Harada14} and strong, widespread shocks \citep[e.g.,][although see also \citealt{YZ13c}, who suggest that a high SiO abundance in CMZ clouds could also be a consequence of a high cosmic ray ionization rate]{JMP97,JMP01,RF04,Mills13}. 

For \meth\, to form via sputtering from shocks requires shocks of velocities sufficient to disrupt grain mantles; as species like \meth\, are relatively loosely bound to these mantles, velocities $\gtrsim10$ \kms$ in continuous or C-type shocks are suggested to be sufficient to enhance the observed \meth\, (and \am) abundances, while shock velocities $>$15 \kms$ will completely release the ice mantles into the gas phase \citep{Caselli97}. For \am\, and \meth\, abundances to both be enhanced is consistent with our observations showing the morphologies of \meth\, and \am ~to be extremely similar on the scales probed here. Further, SMA observations by \cite{Johnston14} illustrate the similar morphologies of \meth\, and SiO, which suggests that the shocks may be yet stronger \citep[shock velocities of 25-40 \kms$ are suggested for CMZ clouds from the abundances of complex molecules and models for their heating][]{JMP01,RF04}. 

The question for \thebrick~then becomes identifying the origin of these shocks. We suggest that shocks due entirely to protostellar outflows are unlikely, given the observed lack of 6 GHz radiatively-excited (Class II) \meth\, masers in this cloud \citep{Caswell96,Caswell10}, which are typically found to be associated with regions of ongoing massive star formation \citep{Voronkov10}. Elsewhere in the Galaxy, Class I masers are observed to be clustered around Class II masers \citep{Slysh94,Valtts00,Ellingsen05,Voronkov14}; the lack of Class II masers in \thebrick~suggests that a different mechanism is responsible for the large \meth\, abundances implied by the Class I masers here. In lieu of protostellar outflows, cloud-cloud collisions have been posited to enhance \meth\, abundances in other CMZ clouds leading to Class I \meth\, masers observed in Sgr A and Sgr B2 \citep{MM97,Sj10}. A cloud collision model has been independently suggested for \thebrick~by \cite{Higuchi14}. However, more recent analyses favor a collapse model for the observed kinematics and morphology of the cloud instead \citep[][Kruijssen et al. in prep]{Rathborne14a,Rathborne15}. Another possibility that should be investigated for this cloud is an interaction with a supernova remnant: the \thebrick~cloud overlaps in projection with a suggested supernova remnant identified by \cite{KF96}. However, it has not yet been demonstrated that this supernova remnant is indeed located at the Galactic center. 

An alternative to a model of shock-enhanced \meth\, abundances in the CMZ is a high cosmic ray ionization rate. In brief: cosmic rays impact the dense gas, exciting the Lyman and Werner bands of H$_2$, generating a weak far-UV field in the cloud interior \citep{Prasad83}. The UV photons are absorbed by molecules in the mantle exterior, which undergo photochemistry (photodissociation, diffusion and recombination), with some of the reaction products with excess energy being desorbed. \cite{YZ13} favor this mechanism over cloud-cloud shocks for the generation of \meth, as they assert that the effects of such shocks are limited to the cloud surfaces, although they do not consider the effects of smaller-scale turbulent shocks in cloud interiors. Given that the distribution of \meth\, masers we observe is clustered and predominantly located in the southern part of the cloud, this might suggest that the masers are indeed limited to locations of large-scale and possibly even surface shocks, rather than more uniformly distributed in the cloud interior (though, as noted above, the apparently nonuniform distribution of \meth\, masers could be an effect of density and/or orientation, and surface shocks could also result in a distribution of masers that, in projection, appears roughly uniform over the entire cloud).

Ultimately, both UV photodesorption from cosmic rays and grain mantle sputtering via shocks appear to be viable mechanisms for generating the observed \meth\, abundances in \thebrick. While our observations of stronger and more numerous masers in the southern part of \thebrick~might slightly favor shocks as the primary mechanism for the abundant \meth\, required to generate these masers, uncertainties as to the role gas density plays in quenching the 36 GHz maser in this cloud make it currently impossible to definitively determine the mechanism responsible. Various future observations could help to distinguish between shocks and cosmic rays as a mechanism for generating the observed \meth\, abundances in the CMZ. First, as previously mentioned, observing other class I \meth\, masers in CMZ clouds (e.g, the 44, 85, and 96 GHz masers) would aid in determining whether all class I maser transitions are stronger and more abundant in the southern part of \thebrick, or if varying density in the cloud selectively quenches the 36 GHz masers. As these are the first observations of extremely abundant 36 GHz maser emission in a giant molecular cloud, it would also be valuable to search for similar emission outside of the CMZ, to determine whether this is truly a phenomenon unique to the inner few hundred parsecs of the Galaxy. While molecular clouds interacting with supernova remnants are not ideal targets as they could be expected to experience both enhanced shocks and cosmic ray ionization, other turbulent clouds lacking star formation should be searched on large-scales for class I \meth\, maser emission. Finally, identifying the heating source for the molecular gas in the CMZ will likely also shed some light on the mechanism responsible for generating the observed abundances of \meth\, and other molecules which are the product of grain-surface chemistry. As both cosmic rays and shocks are suggested to be the most likely source of cloud heating in the CMZ \citep[e.g.,][]{Ao13}, it may be that the mechanism responsible for the heating drives not only the excitation of molecules in this region, but their chemistry as well.

\subsection{Nature of the Continuum Emission}
\label{cmorph}

\subsubsection{The Ionization Source of the Extended Emission}

Much of the continuum emission in \thebrick~(e.g., regions C2, C7 and C9)  is extended on scales of tens of arc seconds (1-2 pc at the Galactic center), forming rough arcs and filaments. We have argued, based on its roughly flat spectral index in several representative regions and its morphological similarity to the continuum emission seen at 90 GHz, that this emission is likely thermal, due to free-free radiation. Assuming this to be the case, the ionization source of this radiation needs to be determined. 

Comparing the spatial distribution of the radio continuum to that of the molecular gas traced by the \amm$\, (3,3) line in Figure \ref{cont_cont}, it is clear that the continuum emission is primarily outlining the eastern edge of the cloud. The spatially extended regions C2, C7 and C9 all lie to the east of the peak of the molecular gas emission, roughly paralleling structures seen in the \amm$\,  line. In particular, C2 exactly parallels a similarly linear \amm$\, feature $5\arcsec$ to the west, while the regions C4, C7, and C8 all seem to trace the outer extent of the \carc feature identified in \amm$\, maps, with a similar spatial offset between the continuum and molecular line emission. C1 may also be a more northern extension of the same structure seen in C2. All together, this structure suggests that the extended emission in \thebrick~is primarily at the cloud's surface, and that its ionization source is external to the cloud and to the east, perhaps a nearby O or B star. There is a known O4-6 supergiant located $\sim$11 pc away in projection to the east of the cloud, at RA = $17^{\mathrm{h}}46^{\mathrm{m}}28.2^{\mathrm{s}}$, Dec = $-28\degr39\arcmin20\arcsec$ \citep{Mauerhan10}. Its bolometric luminosity is estimated to be $10^6$ L$_{\odot}$. Using the parameters of \cite{Martins05}, a supergiant with this luminosity would have a Lyman continuum flux of log Q$_0\sim$49.8. The sum of the Lyman continuum fluxes for the regions in \thebrick~that are catalogued in Table \ref{SItab} is log Q$_0\sim47.8$. Assuming that the cloud intercepts a fraction of the Lyman continuum photons from this sources proportional to its surface area over the distance to the source squared, and that the surface area of \thebrick~as seen by this source is the same as we see in projection (20.5 pc$^2$) then for the cloud to intercept $\sim$1\% of the Lyman continuum photons, the ionizing source would have to be at a distance of $\sim$13 pc, which is consistent with its projected distance, making this a plausible source for ionizing the exterior of \thebrick.

While the filament in C2 appears to have a parallel molecular counterpart, there do not appear to be any molecular counterparts to the linear continuum filament seen in the south of the cloud, next to the shell-shaped C9. This filament is also seen in the 90 GHz maps of \cite{Rathborne14b}. It is then not clear whether this is also an externally-ionized surface feature. A comparison of the continuum emission to the enigmatic HCO$^+$ absorption filaments of \cite{Bally14} does not show any correspondence: none of the HCO$^+$ filaments have a counterpart in either radio continuum or \amm$\, emission. Neither do any of the continuum filaments appear to be oriented parallel to nearby HCO$^+$ filaments. Only the compact continuum source C4 appears to lie along the broad-line absorption filaments, near the junction of filaments 1 and 2, and could be just a chance superposition. However, the \amm$\, emission just to the west of the bottom of the \carc (at RA =17$^{\mathrm{h}}$46$^{\mathrm{m}}$08$^{\mathrm{s}}$, Dec =  -28\degr43\arcmin30\arcsec) does have a steep drop in its brightness, tracing a sharp, nearly linear edge. This edge corresponds to the ``NLA 3" cluster of narrow-line HCO$^+$ absorption features identified by \cite{Bally14}, and it is possible that this edge could represent an analogous absorption feature. However, without the addition of single-dish data to provide a reliable flux zeropoint, it is not clear whether this edge seen in the \amm$\, emission is actually an absorption feature, or simply the absence of emission. 
       
\subsubsection{Evidence for Ongoing Star Formation}

The compact emission sources in and around \thebrick~are of interest for potentially being signatures of the early stages of star formation in this cloud, either ultra- or hypercompact \hii regions. Such regions would be expected to be optically thick, and to have slightly rising spectral indices. Of all of the sources we examine in Section \ref{ion}, three (C3, C4, and C6) appear to have a rising spectrum. Of these, the radio peak of C3 is well aligned with the 90 GHz \citep{Rathborne14b} and 230/280 GHz \citep{Johnston14,Kauffmann13} peaks, while the peak of C4 is slightly spatially offset ($\sim3-5"$) from two adjacent dust continuum peaks detected at 90 and 230 GHz. The stronger of these two peaks corresponds to the location of a previously-identified water maser \citep{Lis94}, marked with a black cross in figure \ref{cont_cont}). Arguments against these radio sources being related to star formation in \thebrick~are that (1) both C3 and C4 (and the more extended C6) are resolved out and not detected in the VLA B-configuration observations of \cite{Rodr13}, suggesting they are not intrinsically compact, and (2) the dust cores associated with C3 and C6 do not show expected signatures of self-gravitation in the column density probability distribution functions (PDFs) of the dust emission constructed by \cite{Johnston14} and \cite{Rathborne14b}. Of course, if the millimeter emission from these sources were instead from optically-thick free free emission instead of thermal dust emission, then any dust column densities inferred for these sources would not be valid. All three sources are assuredly thermal in nature, but whether we are seeing optically-thick free free emission or free-free emission mixed with dust (or a superposition of the two) cannot be determined using available data. Unfortunately, given the weakness of the continuum emission, we do not detect radio recombination line emission, so this cannot be used to determine the contribution of free-free radiation to these fluxes, or to assess whether it is likely to be optically thick. Ultimately, more sensitive observations at both higher and lower radio frequencies are needed to reconstruct the spectral energy distribution of these sources, and to determine their composition.  

The other moderately strong continuum source we detect inside of the cloud, C1, has a flat spectrum consistent with optically-thin free free emission from other nearby, more extended regions of radio continuum emission that appear to trace the external ionization of the cloud, and it is not detected at 230 GHz by \cite{Johnston14}. The two compact continuum sources C5 and C10 are located outside of the bulk of dust and molecular gas in \thebrick, and both have negative spectral indices that suggest their emission is primarily nonthermal. This would make it seem unlikely that they would be associated with star formation in this cloud. However, we do find that one of these sources, C10, is classified as a potential YSO by \cite{An11} based on its infrared spectrum. Although it lies outside of the bulk of \thebrick, it is spatially coincident with the 80 \kms$ cloud. However, without kinematic information from either our continuum observations or the infrared spectrum, it cannot be definitively associated with that cloud. A second YSO candidate identified by \cite{An11} also lies near the 80 \kms$ cloud, however we do not detect a radio counterpart for this source. At present, there is thus no clear evidence from radio continuum observations for advanced stages of star formation in \thebrick.

The other potential signature of early stages of star formation in \thebrick, before the formation of compact HII regions, are the collisionally-excited 36 GHz \meth\, masers. Although the global \meth\, emission in \thebrick~is not likely due to star formation, some of the \meth\, sources are clustered around the continuum sources C5 and C9. However, there are no \meth\, masers near the water maser (and only weak ammonia emission associated with this source). At present, although such correlations are intriguing, there is no way to distinguish masers that could be associated with early stages of star formation and those endemic to the turbulent or cosmic-ray irradiated nature of the cloud.  

\section{Conclusions}

We have detected new weak ($< 1 $mJy) but widespread continuum emission from \thebrick, much of which appears to be due to the external ionization of this cloud by an unknown source. The morphology of the continuum emission includes arcs, filaments, a shell, and multiple compact knots. We have also detected emission from 8 transitions of \amm$, 2 transitions of \cyano, and abundant emission from the 36.2 GHz collisionally-excited maser line of \meth. In total we detect 68 sources whose nonthermal brightness temperatures prove them to be masers, and 81 candidate maser sources, which we expect higher-resolution followup observations will show to be masers as well. Although this is the largest number of these masers ever to be detected in a single molecular cloud, observations of widespread emission in this line throughout the central 200 parsecs suggest that this may be a common feature of Galactic center clouds. As a source of relatively strong and ubiquitous emission, this maser (and potentially other collisionally-excited \meth\, masers), could in the future be a useful tracer of internal cloud kinematics in other turbulent environments, even other Galactic centers.

However, despite these new detections of continuum emission and numerous collisionally-excited \meth\, masers, we find no conclusive evidence for additional star formation in this cloud, apart from the signatures seen by others in a single dust core containing a water maser. We find that several recently-identified compact thermal sources in this cloud which have been suggested to represent embedded star formation actually lie outside of the molecular gas emission, and appear to be mainly associated with more extended structures which we attribute to the external ionization of the cloud. This suggests that \thebrick~truly is unique in the Galactic center, if not the entire Galaxy, as the only massive ($\sim 10^5$ \msun) compact cloud not currently displaying advanced signatures of star formation \citep{Ginsburg12,Tackenberg12,Urquhart14}.  
     
Although observations of radio continuum can rule out the more advanced stages of star formation, in the complicated environment of the Galactic center it is not clear what would represent a ``smoking gun" for early stages of star formation in a cloud. Signatures that are a reliable signpost of the onset of star formation elsewhere in the galaxy: collisionally-excited 36 GHz masers tracing protostellar outflows, hot core chemistry, or elevated temperatures appear to simply be the norm in these clouds \citep[e.g.,][]{YZ13,RT06,Ao13}. Although it may be possible to identify kinematic features such as outflows in a region with simpler kinematics, the extreme turbulent motions of these clouds as a whole make it difficult to ascribe single features to the effects of just one forming star. So, while it appears unlikely that \thebrick~hosts advanced star formation, earlier stages of star formation may still be hidden within. 

\section{Acknowledgements}
We thank Jill Rathborne and Katharine Johnston for sharing their millimeter continuum data and for useful discussions. We also thank Steve Longmore, Diederik Kruijssen, Adam Ginsburg, and the anonymous referee for helpful comments which improved the final version of this paper. We additionally wish to thank the staff of the Karl G. Jansky Very Large Array, and especially to thank Dr. Claire Chandler for assistance with these observations. CCL, NB, and EACM acknowledge support from the NRAO Resident Shared Risk Observing (RSRO) program. EACM also acknowledges support from the NRAO Student Observing Support program. CCL, NB and DAL acknowledge that this material is based upon work supported by the National Science Foundation under Grant No. AST-0907934. 

\bibliographystyle{hapj}
\bibliography{lima}

\clearpage

\section{Figures and Tables}

\clearpage

\begin{table}[ht]
\caption{Observed Fields} 
\centering
\begin{tabular}{ccccc}
\\[0.5ex]
\hline\hline
& & & &  \\
{\bf Field} & {\bf RA} &{\bf Declination} & {\bf Array } &  {\bf  Int.} \\
&  {\bf  (J2000)} & {\bf  (J2000)} & {\bf  Config.} & {\bf  Time}  \\
&  &  &   &  (min) \\
\hline
K-band North  & 17$^{\mathrm{h}}$46$^{\mathrm{m}}$08.95$^{\mathrm{s}}$  & -28\degr 41\arcmin 56.8\arcsec & DnC &24.6  \\
K-band South & 17$^{\mathrm{h}}$46$^{\mathrm{m}}$09.60$^{\mathrm{s}}$  & -28\degr 43\arcmin 24.8\arcsec &  DnC & 24.9 \\
\hline
Ka-band North  & 17$^{\mathrm{h}}$46$^{\mathrm{m}}$08.44$^{\mathrm{s}}$  & -28\degr 42\arcmin 01.0\arcsec &  DnC  & 24.3 \\
Ka-band South  & 17$^{\mathrm{h}}$46$^{\mathrm{m}}$10.26$^{\mathrm{s}}$  & -28\degr 43\arcmin 07.8\arcsec &  DnC  & 24.4  \\
\hline\hline
\end{tabular}
\label{Sources}
\end{table}
\clearpage

\begin{table}[ht]
\caption{Continuum and Spectral Line Image Parameters} 
\centering
\begin{tabular}{cccccccc}
\\[0.5ex]
\hline\hline
\multicolumn{8}{l}{\bf Continuum } \\
\hline
& & & \multicolumn{3}{c}{\bf ------Synthesized Beam------} & & \\
{\bf Band}  & {\bf Center } & {\bf Bandwidth} &{\bf Major} & {\bf Minor} & {\bf Position} & {\bf RMS}  &{\bf Peak  } \\
{\bf } & {\bf Frequency} & & {\bf Axis} &  {\bf Axis} & {\bf Angle} &  {\bf Noise}  & {\bf Intensity} \\
& (GHz) &  (GHz) & (\arcsec) & (\arcsec)  & ($\degr$) &  ($\mu$Jy b$^{-1}$) & (mJy b$^{-1}$) \\
\hline
K (low)  & 24.054 & 0.86 & 2.30 & 1.97 & 70.8 & 47.6 & 1.135  \\
K (high) & 25.375 &  0.86 & 2.19 & 1.87 & 73.7 & 28.6 & 0.811 \\
Ka (low) & 27.515 &  0.86 & 2.21 & 1.87 & 73.7  & 30.0 & 0.636 \\
Ka (high) & 36.350 &  0.86 & 1.59 & 1.51 & 26.2 & 54.5 & 0.672  \\
\hline
\multicolumn{8}{l}{\bf Spectral Line } \\
\hline
& & & \multicolumn{3}{c}{\bf ------Synthesized Beam------} & & \\
{\bf Species+}  & {\bf Rest } & {\bf Channel} &{\bf Major} & {\bf Minor} & {\bf Position} & {\bf RMS}  &{\bf Peak  } \\
{\bf Transition} & {\bf Frequency} & {\bf Width}& {\bf Axis} &  {\bf Axis} & {\bf Angle} &  {\bf per channel}  & {\bf Intensity} \\
& (GHz) &  (\kms$) & (\arcsec) & (\arcsec)  & ($\degr$) &  (mJy b$^{-1}$) & (mJy b$^{-1}$) \\
\hline
\amm$  (1,1) & 23.6945   &  1.58  & 2.83 & 2.58 & 66.9  & 1.6 &  40.2\\
\amm$ (2,2) &  23.7226   &  1.58  & 2.83 & 2.58 & 66.9 & 1.2 & 39.9 \\
\amm$ (3,3) &  23.8701   &  3.14  & 2.81 & 2.56 & 66.8 & 1.9 & 232.4\\
\amm$ (4,4) & 24.1394    &  3.10  & 2.75 & 2.55 & 70.0 & 0.9 & 26.2\\
\amm$ (5,5) & 24.5329    &  3.05  & 2.70 & 2.52 & 69.9 & 0.8 &15.3\\
\amm$ (6,6) & 25.0560    &  2.99   & 2.66 & 2.44 & 71.1 &1.2  & 18.3\\
\amm$ (7,7)  & 25.7152   &  2.91  & 2.59 & 2.37 & 69.1 & 0.7 & 9.3\\      
\cyano$$ (3-2) & 27.2943 &  2.75 & 3.15 & 2.45 & 22.9 & 1.5 & 43.2\\
\amm$ (9,9) &  27.4779    &  2.73 &  2.89 & 2.33 & 45.5  & 1.1 & 11.9  \\
\meth$$  (4-3) &  36.1693 &  1.02 &  2.11   & 1.77 & -2.6  & 2.16 (116)\footnotemark[1]\footnotetext[1]{The larger value is the RMS noise in the channel containing the brightest maser, at v=30 \kms$}  & 28174.0       \\   
\cyano$$  (4-3) &  36.3923 &  2.06 & 1.96 & 1.77 & 15.4 & 3.0 & 40.6 \\
\methcy$$ (2-1)  &  36.7956 &  1.02 & 1.94 & 1.75 & 16.9& 2.6 & 22.5\\     
\hline\hline
\end{tabular}
\label{Imaging}
\end{table}
\clearpage

\begin{table}[ht]
\caption{Continuum Regions} 
\centering
\begin{tabular}{lllllllllrrr}
\\[0.5ex]
\hline\hline
& && \multicolumn{6}{c}{\bf Measured Flux (mJy)\footnotemark[1]\footnotetext[1]{``NA'' indicates this region was outside or near the edge of the field of view}} & \multicolumn{2}{c}{\bf Spectral Index} \\
{\bf }  & {\bf Area} & {\bf Cont.} & {\bf 24.1} & {\bf 25.4}  & {\bf 27.5}  & {\bf 36.4}  & {\bf 90.0}\footnotemark[2]\footnotetext[2]{Values from 3 mm ALMA-only image of \cite{Rathborne14b}} & {\bf 90.0}\footnotemark[3]\footnotetext[3]{Values from single-dish-corrected ALMA image of \cite{Rathborne14b}} & \multicolumn{2}{c}{\bf (24-90 GHz)} & {\bf Log} N$_{Lyc}$ \\
& {\bf (sq$\arcsec$)} & {\bf level} & {\bf GHz} & {\bf GHz} & {\bf GHz}& {\bf GHz}& {\bf GHz}& {\bf GHz}& uncorrected\footnotemark[2] & corrected\footnotemark[3] &  (phot s$^{-1}$)   \\
\hline 
{\bf C1}  & 35.7 &10$\sigma$ &  4.6$\pm$0.1 & 4.6$\pm$0.2 & 4.2$\pm$0.2 & 2.6$\pm$0.1 & 3.1$\pm$0.1 & 6.6$\pm$0.3 & -0.29$\pm0.01$ & 0.27$\pm0.03$  & 46.5  \\ 
{\bf C2}  & 279.1 & 6$\sigma$ & 18.8$\pm$0.1 & 18.7$\pm$0.1 & 15.6$\pm$0.1 & 8.2$\pm$0.2 & 10.5$\pm$0.1 & 27.3$\pm$0.3 & -0.43$\pm$0.01  & 0.28$\pm$0.03  & 47.2 \\
{\bf C3}  & 27.6   & 10$\sigma$ & 2.3$\pm$0.1 & 2.4$\pm$0.1 & 2.1$\pm$0.1 & 2.6$\pm$0.2 & 5.9$\pm$0.1 & 11.3$\pm$0.6 & 0.68$\pm$0.01  & 1.17$\pm$0.05  & 45.9  \\
{\bf C4}  & 14.8   & 6$\sigma$ &  1.0$\pm$0.1 & 0.9$\pm$0.1& 0.9$\pm$0.1 & 0.5$\pm$0.1 & 1.9$\pm$0.1 & 4.2$\pm$0.5 & 0.52$\pm$0.09  & 1.1$\pm$0.1 & 45.9  \\ 
{\bf C5}  & 16.1   & 10$\sigma$ &  6.3$\pm$0.1 & 5.6$\pm$0.2 & 4.0$\pm$0.1 & NA & 1.0$\pm$0.2 & 1.9$\pm$0.4 &-1.31$\pm$0.03 & -0.86$\pm$0.05 &  \ \\
{\bf C6}  & 81.6   & 6$\sigma$ &  4.5$\pm$0.1 & 4.2$\pm$0.1 & 5.5$\pm$0.1 & 6.1$\pm$0.1& 11.8$\pm$0.1 & 26.6$\pm$0.5 & 0.74$\pm$0.07 & 1.34$\pm$0.09 & 46.5  \\
{\bf C7}  & 161.9 & 6$\sigma$ & 10.7$\pm$0.1 & 8.4$\pm$0.2 & 10.1$\pm$0.2 & 5.9$\pm$0.2 & 8.3$\pm$0.1 & 28.4$\pm$0.4 & -0.1$\pm$0.15 & 0.8$\pm$0.18 & 46.9  \\
{\bf C8}  & 164.2 & 6$\sigma$ &  8.4$\pm$0.1 & 5.7$\pm$0.1 & 6.3$\pm$0.1 & 2.8$\pm$0.1 & 6.4$\pm$0.1 & 32.3$\pm$0.5 & -0.1$\pm$0.17 & 1.1$\pm$0.14 & 46.8  \\
{\bf C9}  & 521.8 & 6$\sigma$ & 43.4$\pm$0.1 & 34.1$\pm$0.2 & 37.7$\pm$0.2 & 35.8$\pm$0.2& 24.4$\pm$0.3 & 80.9$\pm$0.3 & -0.3$\pm$0.25 & 0.6$\pm$0.3 & 47.5  \\
{\bf C10}& 7.7   & 10$\sigma$ &   1.6$\pm$0.1 & 1.6$\pm$0.2 & NA & NA & 0.6$\pm$0.1 & 1.0$\pm$0.1 & -0.73$\pm$0.03 & -0.35$\pm$0.01 &  \\
\hline\hline
\end{tabular}
\label{SItab}
\end{table}
\clearpage

\LongTables
\centering
\begin{deluxetable*}{cccccccccc}
\tablecaption{\\ \sc{36 GHz \meth~ Masers}}
\tablecomments{This is Table 4: \sc{36 GHz \meth~ Masers}}
\\[0.5ex]
\hline\hline
& & & & & & & & & \\[0.05cm]
{\bf ID} & {\bf Maser   Name}&  {\bf RA}&       {\bf Dec}&      {\bf Velocity}& {\bf FWHM}&     {\bf I$_{\mathrm{peak}}$}&        {\bf Flux}&    {\bf T$_{\mathrm{B}}$}&       {\bf Resolved?} \\[0.05cm]
& & HH:MM:SS.s   & DD:MM:SS.s &         \kms$ &  \kms$     & Jy beam$^{-1}$ &    Jy \kms$ & K      & \\[0.05cm]
\hline
& & & & & & & & \\[0.05cm]
M1&    G0.2398299+0.0034302   &17:46:10.62   &-28:43:46.6   & 30.447   & 3.184   & 76.212   &333.646&    19269   &YES \\[0.05cm]
M2&    G0.2352551+0.0058162   &17:46:09.41   &-28:43:56.2   & 38.735   & 3.484   & 32.446   &152.436&     8203   &YES \\[0.05cm]
M3&    G0.2678706+0.0317273   &17:46:08.00   &-28:41:27.4   & 30.447   & 1.836   & 27.686   & 76.003&     7000   & NO \\[0.05cm]
M4&    G0.2438565+0.0040277   &17:46:11.06   &-28:43:33.1   & 35.627   & 2.912   & 21.630   & 85.486&     5468   & NO \\[0.05cm]
M5&    G0.2349796+0.0064290   &17:46:09.23   &-28:43:55.9   & 40.807   & 4.172   & 20.860   &148.052&     5274   &YES \\[0.05cm]
M6&    G0.2431785+0.0049810   &17:46:10.74   &-28:43:33.4   & 21.124   & 2.981   & 20.305   & 57.793&     5133   & NO \\[0.05cm]
M7&    G0.2411601+0.0076552   &17:46:09.82   &-28:43:34.6   & 33.555   & 4.162   & 19.237   & 60.459&     4863   & NO \\[0.05cm]
M8&    G0.2351776+0.0067447   &17:46:09.19   &-28:43:54.7   & 39.771   & 3.814   & 17.971   &130.015&     4543   &YES \\[0.05cm]
M9&    G0.2693832+0.0342102   &17:46:07.64   &-28:41:18.1   &  6.620   & 6.638   & 14.104   & 86.632&     3565   & NO \\[0.05cm]
M10&   G0.2584673+0.0229753   &17:46:08.71   &-28:42:12.7   & 31.483   & 2.249   & 10.252   & 29.481&     2592   & NO \\[0.05cm]
M11&   G0.2357657+0.0064198   &17:46:09.35   &-28:43:53.5   & 35.627   & 5.246   & 10.178   & 79.186&     2573   & NO \\[0.05cm]
M12&   G0.2584886+0.0240620   &17:46:08.46   &-28:42:10.6   & 39.771   & 4.477   &  9.906   & 53.927&     2504   &YES \\[0.05cm]
M13&   G0.2359542+0.0081933   &17:46:08.96   &-28:43:49.6   & 36.663   & 3.958   &  9.344   & 63.827&     2362   &YES \\[0.05cm]
M14&   G0.2446676+0.0023749   &17:46:11.56   &-28:43:33.7   & 27.339   & 2.525   &  8.720   & 27.893&     2204   & NO \\[0.05cm]
M15&   G0.2353505+0.0087040   &17:46:08.75   &-28:43:50.5   & 38.735   & 3.241   &  8.510   & 52.511&     2151   &YES \\[0.05cm]
M16&   G0.2405942+0.0045786   &17:46:10.46   &-28:43:42.1   & 29.411   & 3.563   &  8.304   & 29.414&     2099   &YES \\[0.05cm]
M17&   G0.2356106+0.0082769   &17:46:08.89   &-28:43:50.5   & 39.771   & 3.778   &  8.233   & 45.093&     2081   & NO \\[0.05cm]
M18&   G0.2549814+0.0248543   &17:46:07.77   &-28:42:19.9   & 19.052   & 5.856   &  7.682   & 51.912&     1942   &YES \\[0.05cm]
M19&   G0.2335835+0.0079207   &17:46:08.68   &-28:43:57.4   & 42.879   & 3.324   &  6.400   & 38.649&     1618   & NO \\[0.05cm]
M20&   G0.2440175+0.0034427   &17:46:11.27   &-28:43:33.7   & 33.555   & 5.959   &  6.225   & 36.388&     1573   & NO \\[0.05cm]
M21&   G0.2396891+0.0118337   &17:46:08.64   &-28:43:31.3   & 49.095   & 4.186   &  6.195   & 21.655&     1566   & NO \\[0.05cm]
M22&   G0.2444911+0.0031456   &17:46:11.35   &-28:43:32.8   & 33.555   & 3.669   &  6.150   & 32.977&     1554   & NO \\[0.05cm]
M23&   G0.2454012+0.0016506   &17:46:11.83   &-28:43:32.8   & 31.483   & 2.822   &  6.107   & 24.441&     1543   &YES \\[0.05cm]
M24&   G0.2436696+0.0107445   &17:46:09.46   &-28:43:21.1   & 49.095   & 2.670   &  5.156   & 18.498&     1303   & NO \\[0.05cm]
M25&   G0.2572398+0.0184216   &17:46:09.60   &-28:42:25.0   &  7.656   & 5.076   &  5.124   & 33.197&     1295   &YES \\[0.05cm]
M26&   G0.2418468+0.0097322   &17:46:09.44   &-28:43:28.6   & 44.951   & 2.828   &  4.877   & 16.039&     1233   & NO \\[0.05cm]
M27&   G0.2423057+0.0051327   &17:46:10.58   &-28:43:35.8   & 31.483   & 6.288   &  4.842   & 42.715&     1224   &YES \\[0.05cm]
M28&   G0.2422560+0.0061758   &17:46:10.33   &-28:43:34.0   & 35.627   & 5.536   &  4.668   & 42.038&     1180   &YES \\[0.05cm]
M29&   G0.2397862+0.0057454   &17:46:10.08   &-28:43:42.4   & 27.339   & 3.447   &  4.482   & 19.467&     1133   &YES \\[0.05cm]
M30&   G0.2429928+0.0049656   &17:46:10.71   &-28:43:34.0   & 25.267   & 5.322   &  4.395   & 27.276&     1111   &YES \\[0.05cm]
M31&   G0.2716946+0.0284917   &17:46:09.30   &-28:41:21.7   & -2.704   & 2.087   &  4.354   & 13.012&     1100   & NO \\[0.05cm]
M32&   G0.2587389+0.0250928   &17:46:08.25   &-28:42:07.9   & 34.591   & 3.461   &  4.336   & 27.130&     1096   &YES \\[0.05cm]
M33&   G0.2411665+0.0066833   &17:46:10.05   &-28:43:36.4   & 40.807   & 5.608   &  4.201   & 67.957&     1062   &YES \\[0.05cm]
M34&   G0.2556408+0.0245727   &17:46:07.93   &-28:42:18.4   & 39.771   & 7.755   &  3.710   & 29.827&      937   &YES \\[0.05cm]
M35&   G0.2376346+0.0091190   &17:46:08.98   &-28:43:42.7   & 37.699   & 1.279   &  3.699   &  7.069&      935   & NO \\[0.05cm]
M36&   G0.2424071+0.0092927   &17:46:09.62   &-28:43:27.7   & 47.023   & 3.272   &  3.621   & 20.174&      915   &YES \\[0.05cm]
M37&   G0.2397982+0.0082897   &17:46:09.48   &-28:43:37.6   & 43.915   & 4.043   &  3.486   & 33.611&      881   &YES \\[0.05cm]
M38&   G0.2495436+0.0129548   &17:46:09.78   &-28:42:58.9   & 19.052   & 3.634   &  3.421   & 21.531&      864   &YES \\[0.05cm]
M39&   G0.2447480+0.0032044   &17:46:11.38   &-28:43:31.9   & 30.447   & 4.930   &  3.377   & 20.993&      853   &YES \\[0.05cm]
M40&   G0.2664108+0.0362072   &17:46:06.75   &-28:41:23.5   &  3.512   & 2.686   &  2.898   & 11.934&      732   &YES \\[0.05cm]
M41&   G0.2443042+0.0098623   &17:46:09.76   &-28:43:20.8   & 50.131   & 1.878   &  2.850   &  6.041&      720   &YES \\[0.05cm]
M42&   G0.2544308+0.0238358   &17:46:07.93   &-28:42:23.5   & 42.879   & 4.374   &  2.814   & 22.516&      711   &YES \\[0.05cm]
M43&   G0.2671146+0.0336096   &17:46:07.45   &-28:41:26.2   &  9.728   & 1.446   &  2.714   &  6.569&      686   & NO \\[0.05cm]
M44&   G0.2470599+0.0037337   &17:46:11.58   &-28:43:23.8   & 35.627   & 2.917   &  2.650   & 10.509&      670   &YES \\[0.05cm]
M45&   G0.2396685+0.0062592   &17:46:09.94   &-28:43:41.8   & 33.555   & 3.344   &  2.578   & 17.679&      651   &YES \\[0.05cm]
M46&   G0.2361667+0.0128114   &17:46:07.91   &-28:43:40.3   & 48.059   & 1.179   &  2.549   &  4.716&      644   & NO \\[0.05cm]
M47&   G0.2592935+0.0233809   &17:46:08.73   &-28:42:09.4   & 41.843   & 2.275   &  2.392   &  6.557&      604   & NO \\[0.05cm]
M48&   G0.2496394+0.0135987   &17:46:09.64   &-28:42:57.4   & 35.627   & 1.599   &  2.382   &  4.899&      602   & NO \\[0.05cm]
M49&   G0.2588943+0.0232354   &17:46:08.71   &-28:42:10.9   & 40.807   & 3.613   &  2.340   & 12.201&      591   & NO \\[0.05cm]
M50&   G0.2667441+0.0313341   &17:46:07.93   &-28:41:31.6   & 31.483   & 1.821   &  2.328   &  4.342&      588   & NO \\[0.05cm]
M51&   G0.2414639+0.0049129   &17:46:10.51   &-28:43:38.8   & 33.555   & 5.188   &  2.203   & 14.952&      556   & NO \\[0.05cm]
M52&   G0.2587707+0.0224769   &17:46:08.87   &-28:42:12.7   & 39.771   & 3.374   &  2.190   &  7.639&      553   & NO \\[0.05cm]
M53&   G0.2597762+0.0238700   &17:46:08.68   &-28:42:07.0   & 39.771   & 2.907   &  2.180   &  6.395&      551   & NO \\[0.05cm]
M54&   G0.2465863+0.0040308   &17:46:11.44   &-28:43:24.7   & 39.771   & 5.857   &  2.137   & 10.270&      540   & NO \\[0.05cm]
M55&   G0.2467184+0.0100636   &17:46:10.05   &-28:43:13.0   & 33.555   & 2.486   &  2.110   & 13.584&      533   &YES \\[0.05cm]
M56&   G0.2392155+0.0121308   &17:46:08.50   &-28:43:32.2   & 50.131   & 5.607   &  2.022   & 14.468&      511   & NO \\[0.05cm]
M57&   G0.2432270+0.0106702   &17:46:09.41   &-28:43:22.6   & 35.627   & 2.587   &  2.016   & 10.836&      509   &YES \\[0.05cm]
M58&   G0.2568865+0.0199632   &17:46:09.19   &-28:42:23.2   &  9.728   & 2.790   &  2.016   &  7.985&      509   & NO \\[0.05cm]
M59&   G0.2411171+0.0054824   &17:46:10.33   &-28:43:38.8   & 38.735   & 5.613   &  1.980   & 25.949&      500   &YES \\[0.05cm]
M60&   G0.2444434+0.0104350   &17:46:09.64   &-28:43:19.3   & 36.663   & 3.288   &  1.977   & 16.399&      499   &YES \\[0.05cm]
M61&   G0.2439377+0.0155918   &17:46:08.37   &-28:43:11.2   & 37.699   & 1.277   &  1.958   &  2.131&      495   & NO \\[0.05cm]
M62&   G0.2664301+0.0163067   &17:46:11.40   &-28:42:00.7   & -4.776   & 1.256   &  1.871   &  2.169&      473   & NO \\[0.05cm]
M63&   G0.2541338+0.0233622   &17:46:08.00   &-28:42:25.3   & 47.023   & 4.585   &  1.859   &  7.060&      470   & NO \\[0.05cm]
M64&   G0.2463291+0.0062161   &17:46:10.90   &-28:43:21.4   & 43.915   & 1.510   &  1.802   &  4.377&      455   & NO \\[0.05cm]
M65&   G0.2411450+0.0055969   &17:46:10.30   &-28:43:38.5   & 43.915   & 2.842   &  1.728   &  9.365&      436   & NO \\[0.05cm]
M66&   G0.2411760+0.0052255   &17:46:10.40   &-28:43:39.1   & 27.339   & 4.334   &  1.682   &  6.055&      425   & NO \\[0.05cm]
M67&   G0.2416611+0.0097167   &17:46:09.41   &-28:43:29.2   & 49.095   & 3.599   &  1.665   &  7.314&      420   & NO \\[0.05cm]
M68&   G0.2550027+0.0259409   &17:46:07.52   &-28:42:17.8   & 32.519   & 2.458   &  1.640   &  6.657&      414   & NO \\[0.05cm]
\hline
\label{Masers}
\end{deluxetable*}
\clearpage
\clearpage

\LongTables
\centering
\begin{deluxetable*}{cccccccccc}
\tablecaption{\\ \sc {36 GHz \meth~ Maser Candidates}}
\tablecomments{This is Table 4: \sc{36 GHz \meth~ Maser Candidates}}
\\[0.5ex]
\hline\hline
& & & & & & & & & \\[0.05cm]
{\bf ID} & {\bf Candidate  Name}&  {\bf RA}&       {\bf Dec}&      {\bf Velocity}& {\bf FWHM}&     {\bf I$_{\mathrm{peak}}$}&        {\bf Flux}&    {\bf T$_{\mathrm{B}}$}&       {\bf Resolved?} \\[0.05cm]
& & HH:MM:SS.s   & DD:MM:SS.s &         \kms$ &  \kms$     & Jy beam$^{-1}$ &    Jy \kms$ & K      & \\[0.05cm]
\hline
& & & & & & & & \\[0.05cm]
CM1&    G0.2684202+0.0349905   &17:46:07.32   &-28:41:19.6   & 27.339   & 1.126   &  1.565   &  1.958&      395   & NO \\[0.05cm]
CM2&    G0.2365820+0.0105272   &17:46:08.50   &-28:43:43.3   & 42.879   & 1.494   &  1.555   &  7.302&      393   &YES \\[0.05cm]
CM3&    G0.2726984+0.0258817   &17:46:10.05   &-28:41:23.5   &  1.440   & 1.211   &  1.472   &  2.500&      372   & NO \\[0.05cm]
CM4&    G0.2414813+0.0109734   &17:46:09.10   &-28:43:27.4   & 40.807   & 5.521   &  1.450   & 13.575&      366   &YES \\[0.05cm]
CM5&    G0.2396343+0.0071166   &17:46:09.73   &-28:43:40.3   & 35.627   & 8.515   &  1.433   & 17.606&      362   &YES \\[0.05cm]
CM6&    G0.2373767+0.0135482   &17:46:07.91   &-28:43:35.2   & 55.311   & 4.992   &  1.332   &  7.914&      336   &YES \\[0.05cm]
CM7&    G0.2639542+0.0338325   &17:46:06.95   &-28:41:35.5   & 25.267   & 1.212   &  1.274   &  2.257&      322   & NO \\[0.05cm]
CM8&    G0.2574507+0.0167933   &17:46:10.01   &-28:42:27.4   &  7.656   & 4.640   &  1.262   &  5.075&      318   & NO \\[0.05cm]
CM9&    G0.2563183+0.0151279   &17:46:10.24   &-28:42:34.0   & 14.908   & 5.059   &  1.248   & 18.386&      315   &YES \\[0.05cm]
CM10&   G0.2716086+0.0263896   &17:46:09.78   &-28:41:25.9   & -0.632   & 4.520   &  1.147   & 10.161&      289   & NO \\[0.05cm]
CM11&   G0.2671442+0.0292350   &17:46:08.48   &-28:41:34.3   & 12.836   & 4.286   &  1.139   &  6.661&      287   & NO \\[0.05cm]
CM12&   G0.2391512+0.0088715   &17:46:09.25   &-28:43:38.5   & 36.663   & 1.063   &  1.117   &  1.160&      282   & NO \\[0.05cm]
CM13&   G0.2409922+0.0114562   &17:46:08.91   &-28:43:28.0   & 49.095   & 3.557   &  1.104   &  6.213&      279   &YES \\[0.05cm]
CM14&   G0.2572715+0.0158058   &17:46:10.21   &-28:42:29.8   & 16.980   & 4.448   &  1.074   &  9.304&      271   &YES \\[0.05cm]
CM15&   G0.2568652+0.0188767   &17:46:09.44   &-28:42:25.3   & 10.764   & 5.441   &  1.065   &  9.814&      269   &YES \\[0.05cm]
CM16&   G0.2574106+0.0163785   &17:46:10.10   &-28:42:28.3   & 11.800   & 4.718   &  1.050   &  6.166&      265   & NO \\[0.05cm]
CM17&   G0.2403644+0.0091223   &17:46:09.37   &-28:43:34.3   & 33.555   & $<$1.036 &  1.020   &  0.928&      257   & NO \\[0.05cm]
CM18&   G0.2580051+0.0258172   &17:46:07.98   &-28:42:08.8   & 25.267   & 4.792   &  0.993   &  4.854&      251   &YES \\[0.05cm]
CM19&   G0.2460093+0.0113884   &17:46:09.64   &-28:43:12.7   & 35.627   & 2.245   &  0.992   &  3.101&      250   &YES \\[0.05cm]
CM20&   G0.2685456+0.0312597   &17:46:08.21   &-28:41:26.2   &  9.728   & 2.218   &  0.934   &  5.817&      236   &YES \\[0.05cm]
CM21&   G0.2540684+0.0138556   &17:46:10.21   &-28:42:43.3   & 16.980   & 4.247   &  0.895   & 12.616&      226   &YES \\[0.05cm]
CM22&   G0.2623100+0.0275602   &17:46:08.18   &-28:41:52.3   & 28.375   & 1.281   &  0.883   &  1.387&      223   & NO \\[0.05cm]
CM23&   G0.2720112+0.0260490   &17:46:09.92   &-28:41:25.3   &  2.476   & 5.335   &  0.877   &  3.608&      221   & NO \\[0.05cm]
CM24&   G0.2677259+0.0298821   &17:46:08.41   &-28:41:31.3   &  5.584   & 6.154   &  0.853   &  3.208&      215   & NO \\[0.05cm]
CM25&   G0.2381058+0.0048198   &17:46:10.05   &-28:43:49.3   & 27.339   & 1.216   &  0.811   &  1.138&      205   & NO \\[0.05cm]
CM26&   G0.2450510+0.0049501   &17:46:11.01   &-28:43:27.7   & 23.195   & 3.486   &  0.781   &  5.021&      197   &YES \\[0.05cm]
CM27&   G0.2591797+0.0211644   &17:46:09.23   &-28:42:13.9   & 43.915   & 2.376   &  0.776   &  3.477&      196   & NO \\[0.05cm]
CM28&   G0.2566710+0.0158306   &17:46:10.12   &-28:42:31.6   & 11.800   & 4.636   &  0.776   &  6.625&      196   & NO \\[0.05cm]
CM29&   G0.2580207+0.0148956   &17:46:10.53   &-28:42:29.2   & 34.591   & 1.210   &  0.751   &  0.889&      189   & NO \\[0.05cm]
CM30&   G0.2576131+0.0224553   &17:46:08.71   &-28:42:16.3   & 47.023   & 1.371   &  0.746   &  1.814&      188   & NO \\[0.05cm]
CM31&   G0.2451130+0.0042073   &17:46:11.19   &-28:43:28.9   & 28.375   & 2.262   &  0.743   &  2.622&      187   &YES \\[0.05cm]
CM32&   G0.2382249+0.0127961   &17:46:08.21   &-28:43:34.0   & 56.347   & 5.856   &  0.734   &  7.584&      185   &YES \\[0.05cm]
CM33&   G0.2488573+0.0086336   &17:46:10.69   &-28:43:09.1   & 35.627   & 1.670   &  0.733   &  1.084&      185   & NO \\[0.05cm]
CM34&   G0.2479380+0.0093424   &17:46:10.40   &-28:43:10.6   & 36.663   & 2.849   &  0.713   &  1.683&      180   & NO \\[0.05cm]
CM35&   G0.2632957+0.0318697   &17:46:07.32   &-28:41:41.2   & 27.339   & 1.717   &  0.700   &  2.202&      176   & NO \\[0.05cm]
CM36&   G0.2634060+0.0345724   &17:46:06.70   &-28:41:35.8   &  1.440   & 3.715   &  0.700   &  3.895&      176   &YES \\[0.05cm]
CM37&   G0.2399226+0.0045600   &17:46:10.38   &-28:43:44.2   & 27.339   & 3.211   &  0.696   &  3.186&      175   &YES \\[0.05cm]
CM38&   G0.2434175+0.0164461   &17:46:08.09   &-28:43:11.2   & 45.987   & 1.161   &  0.690   &  1.192&      174   & NO \\[0.05cm]
CM39&   G0.2645475+0.0285323   &17:46:08.27   &-28:41:43.6   & 33.555   & 1.549   &  0.688   &  1.715&      173   & NO \\[0.05cm]
CM40&   G0.2474365+0.0095251   &17:46:10.28   &-28:43:11.8   & 33.555   & 1.863   &  0.662   &  1.773&      167   & NO \\[0.05cm]
CM41&   G0.2584992+0.0288511   &17:46:07.34   &-28:42:01.6   &  8.692   & 1.147   &  0.658   &  0.697&      166   & NO \\[0.05cm]
CM42&   G0.2576803+0.0252290   &17:46:08.07   &-28:42:10.9   & 16.980   & 6.136   &  0.647   &  6.469&      163   &YES \\[0.05cm]
CM43&   G0.2453628+0.0097262   &17:46:09.94   &-28:43:17.8   & 34.591   & 1.174   &  0.626   &  0.809&      158   & NO \\[0.05cm]
CM44&   G0.2454910+0.0010222   &17:46:11.99   &-28:43:33.7   & 83.283   & 4.220   &  0.611   &  4.026&      154   &YES \\[0.05cm]
CM45&   G0.2561685+0.0052779   &17:46:12.52   &-28:42:52.9   & 42.879   & 1.074   &  0.607   &  0.550&      153   & NO \\[0.05cm]
CM46&   G0.2412376+0.0067267   &17:46:10.05   &-28:43:36.1   & 25.267   & 1.583   &  0.594   &  1.150&      150   & NO \\[0.05cm]
CM47&   G0.2673045+0.0308945   &17:46:08.11   &-28:41:30.7   &  9.728   & 1.799   &  0.592   &  3.120&      149   &YES \\[0.05cm]
CM48&   G0.2718999+0.0255908   &17:46:10.01   &-28:41:26.5   & -6.848   & 3.647   &  0.570   &  3.638&      144   &YES \\[0.05cm]
CM49&   G0.2540717+0.0241052   &17:46:07.82   &-28:42:24.1   & 24.231   & 4.852   &  0.556   &  4.864&      140   & NO \\[0.05cm]
CM50&   G0.2379498+0.0111649   &17:46:08.55   &-28:43:37.9   & 45.987   & 1.641   &  0.553   &  1.241&      139   &YES \\[0.05cm]
CM51&   G0.2564375+0.0208610   &17:46:08.91   &-28:42:22.9   &  7.656   & 3.531   &  0.545   &  2.843&      137   & NO \\[0.05cm]
CM52&   G0.2396789+0.0155356   &17:46:07.77   &-28:43:24.4   & 15.944   & 1.258   &  0.544   &  0.910&      137   & NO \\[0.05cm]
CM53&   G0.2672072+0.0262474   &17:46:09.19   &-28:41:39.7   & 11.800   & 2.678   &  0.529   &  5.735&      133   &YES \\[0.05cm]
CM54&   G0.2591622+0.0173474   &17:46:10.12   &-28:42:21.1   & 12.836   & 2.141   &  0.511   &  1.560&      129   & NO \\[0.05cm]
CM55&   G0.2668194+0.0286468   &17:46:08.57   &-28:41:36.4   & 13.872   & 4.400   &  0.490   &  3.069&      124   & NO \\[0.05cm]
CM56&   G0.2655322+0.0350864   &17:46:06.88   &-28:41:28.3   &  3.512   & 1.767   &  0.488   &  0.968&      123   &YES \\[0.05cm]
CM57&   G0.2391662+0.0109298   &17:46:08.78   &-28:43:34.6   & 52.203   & 3.941   &  0.469   &  2.405&      118   &YES \\[0.05cm]
CM58&   G0.2633546+0.0316127   &17:46:07.38   &-28:41:41.5   &  1.440   & 1.208   &  0.462   &  0.802&      116   & NO \\[0.05cm]
CM59&   G0.2565054+0.0213903   &17:46:08.80   &-28:42:21.7   & 11.800   & 2.355   &  0.450   &  1.856&      113   & NO \\[0.05cm]
CM60&   G0.2621404+0.0336065   &17:46:06.75   &-28:41:41.5   & 25.267   & 1.464   &  0.441   &  1.561&      111   &YES \\[0.05cm]
CM61&   G0.2523601+0.0128155   &17:46:10.21   &-28:42:50.5   & 12.836   & 3.974   &  0.409   &  2.917&      103   &YES \\[0.05cm]
CM62&   G0.2472548+0.0045354   &17:46:11.42   &-28:43:21.7   & 24.231   & 2.302   &  0.407   &  1.038&      102   & NO \\[0.05cm]
CM63&   G0.2536039+0.0149391   &17:46:09.89   &-28:42:42.7   & 12.836   & 3.988   &  0.405   &  4.935&      102   &YES \\[0.05cm]
CM64&   G0.2490084+0.0117507   &17:46:09.98   &-28:43:02.8   & 18.016   & 3.543   &  0.398   &  4.842&      100   &YES \\[0.05cm]
CM65&   G0.2501440+0.0129301   &17:46:09.87   &-28:42:57.1   & 14.908   & 4.048   &  0.396   &  2.308&      100   & NO \\[0.05cm]
CM66&   G0.2586394+0.0164434   &17:46:10.26   &-28:42:24.4   & 15.944   & 3.890   &  0.391   &  2.884&       98   &YES \\[0.05cm]
CM67&   G0.2407096+0.0152850   &17:46:07.98   &-28:43:21.7   & 16.980   & 1.754   &  0.389   &  1.019&       98   & NO \\[0.05cm]
CM68&   G0.2501568+0.0109861   &17:46:10.33   &-28:43:00.7   & 24.231   & 2.672   &  0.369   &  1.039&       93   & NO \\[0.05cm]
CM69&   G0.2637776+0.0261113   &17:46:08.73   &-28:41:50.5   & 11.800   & $<$1.036 &  0.366   &  0.329&       92   & NO \\[0.05cm]
CM70&   G0.2437590+0.0123602   &17:46:09.10   &-28:43:17.8   & 21.124   & 3.610   &  0.364   &  0.695&       91   & NO \\[0.05cm]
CM71&   G0.2537285+0.0112090   &17:46:10.78   &-28:42:49.3   & 22.159   & 1.804   &  0.362   &  0.833&       91   & NO \\[0.05cm]
CM72&   G0.2556194+0.0127505   &17:46:10.69   &-28:42:40.6   & 21.124   & 3.338   &  0.360   &  1.690&       91   &YES \\[0.05cm]
CM73&   G0.2439291+0.0125614   &17:46:09.07   &-28:43:16.9   & 25.267   & 2.348   &  0.359   &  1.394&       90   & NO \\[0.05cm]
CM74&   G0.2503013+0.0150752   &17:46:09.39   &-28:42:52.6   & 22.159   & $<$1.036 &  0.346   &  0.257&       87   & NO \\[0.05cm]
CM75&   G0.2699078+0.0306250   &17:46:08.55   &-28:41:23.2   & -7.884   & 5.485   &  0.341   & 15.765&       86   &YES \\[0.05cm]
CM76&   G0.2562854+0.0222324   &17:46:08.57   &-28:42:20.8   & 11.800   & 1.849   &  0.340   &  1.042&       85   & NO \\[0.05cm]
CM77&   G0.2510821+0.0115495   &17:46:10.33   &-28:42:56.8   & 27.339   & 2.811   &  0.337   &  1.618&       85   & NO \\[0.05cm]
CM78&   G0.2479022-0.0005347   &17:46:12.70   &-28:43:29.2   & 81.211   & 6.053   &  0.330   &  9.609&       83   &YES \\[0.05cm]
CM79&   G0.2527749+0.0127753   &17:46:10.28   &-28:42:49.3   & 16.980   & 3.675   &  0.312   &  2.886&       78   &YES \\[0.05cm]
CM80&   G0.2613076+0.0341730   &17:46:06.49   &-28:41:43.0   &  0.404   & 1.182   &  0.303   &  0.462&       76   & NO \\[0.05cm]
\hline

\label{Candidates}
\end{deluxetable*}
\clearpage

\renewcommand\thefigure{7}
\begin{figure}
\includegraphics[scale=0.85]{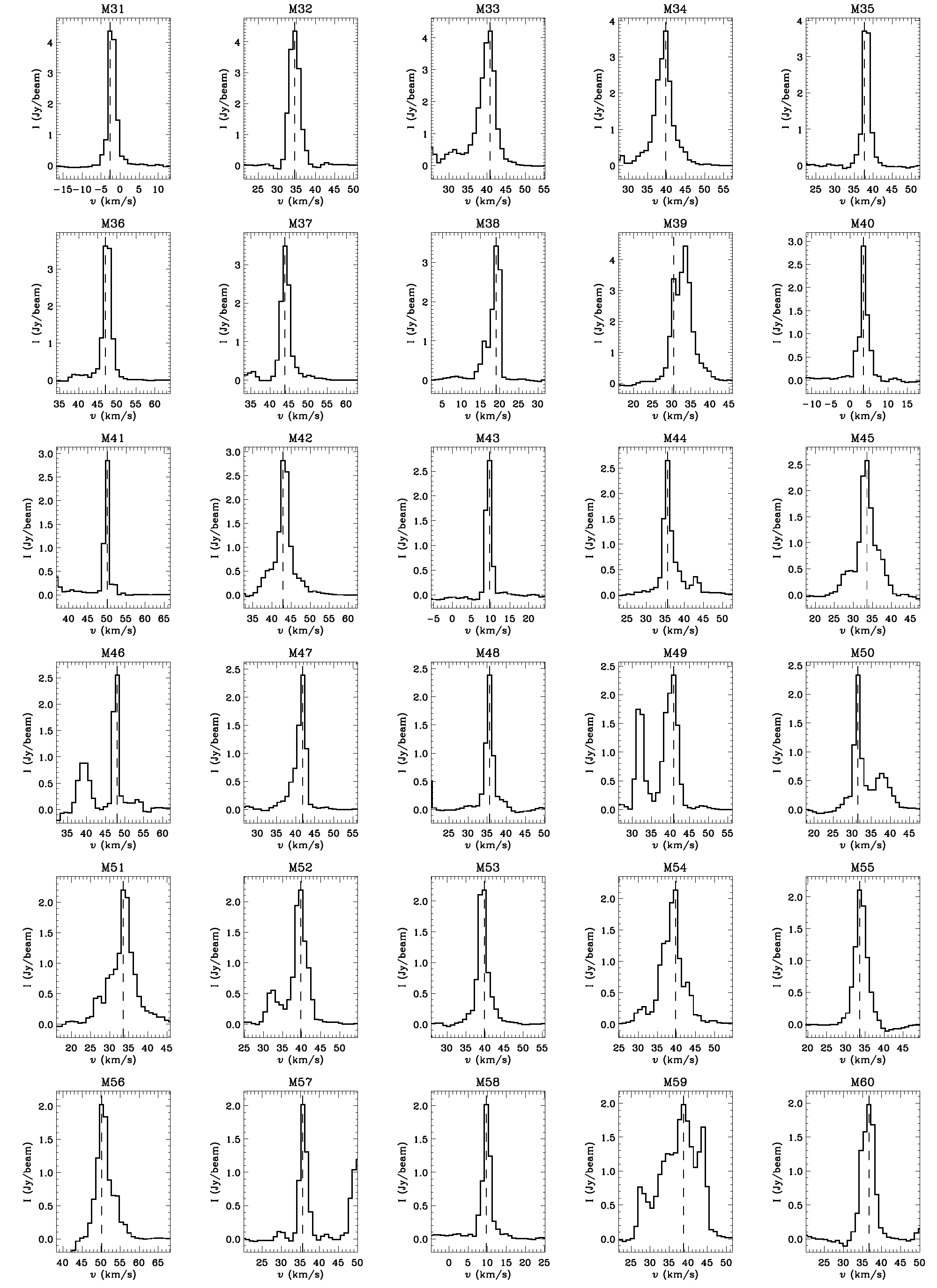}
\caption{ Maser spectra, continued.}
\label{maserspect2}
\end{figure}

\renewcommand\thefigure{7}
\begin{figure}
\includegraphics[scale=0.85]{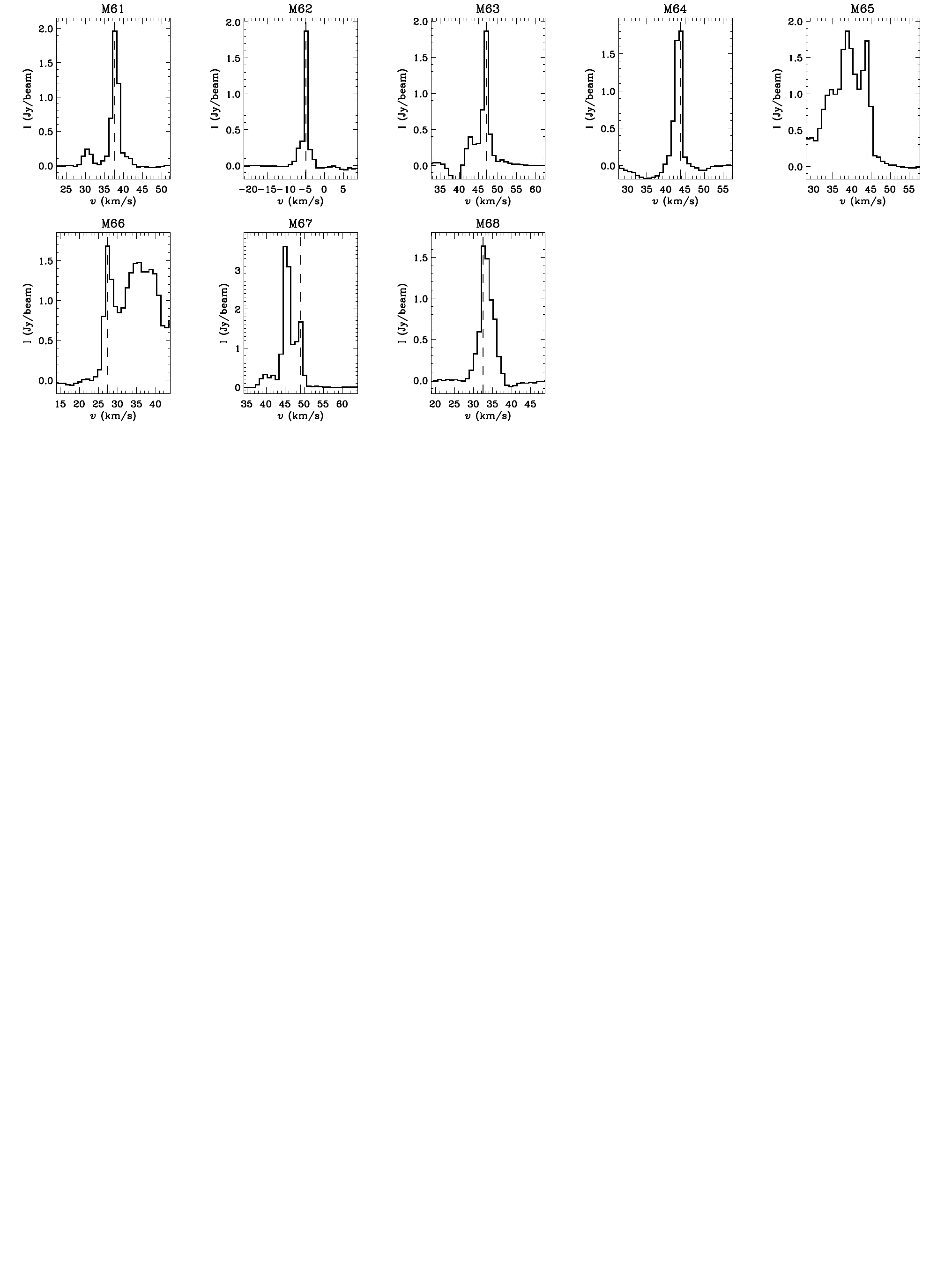}
\caption{ Maser spectra, continued.}
\label{maserspect3}
\end{figure}

\renewcommand\thefigure{8}
\begin{figure}
\includegraphics[scale=0.85]{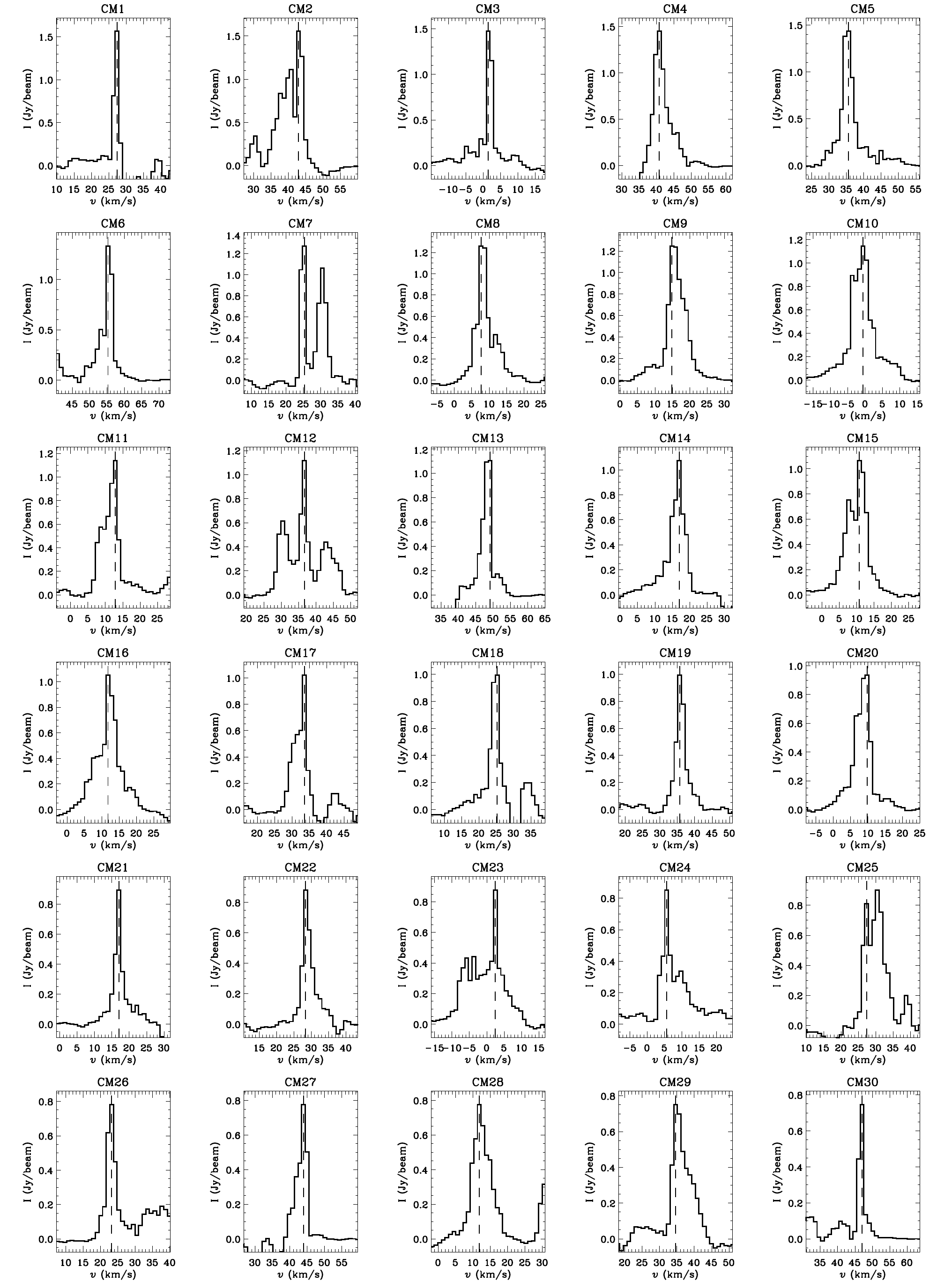}
\caption{ Spectra of all catalogued candidate masers. Each spectrum corresponds to the pixel associated with the peak emission in the source defined by {\it Clumpfind}. The conversion factor to go from Jy to K is 250.5}
\label{candspect1}
\end{figure}

\renewcommand\thefigure{8}
\begin{figure}
\includegraphics[scale=0.85]{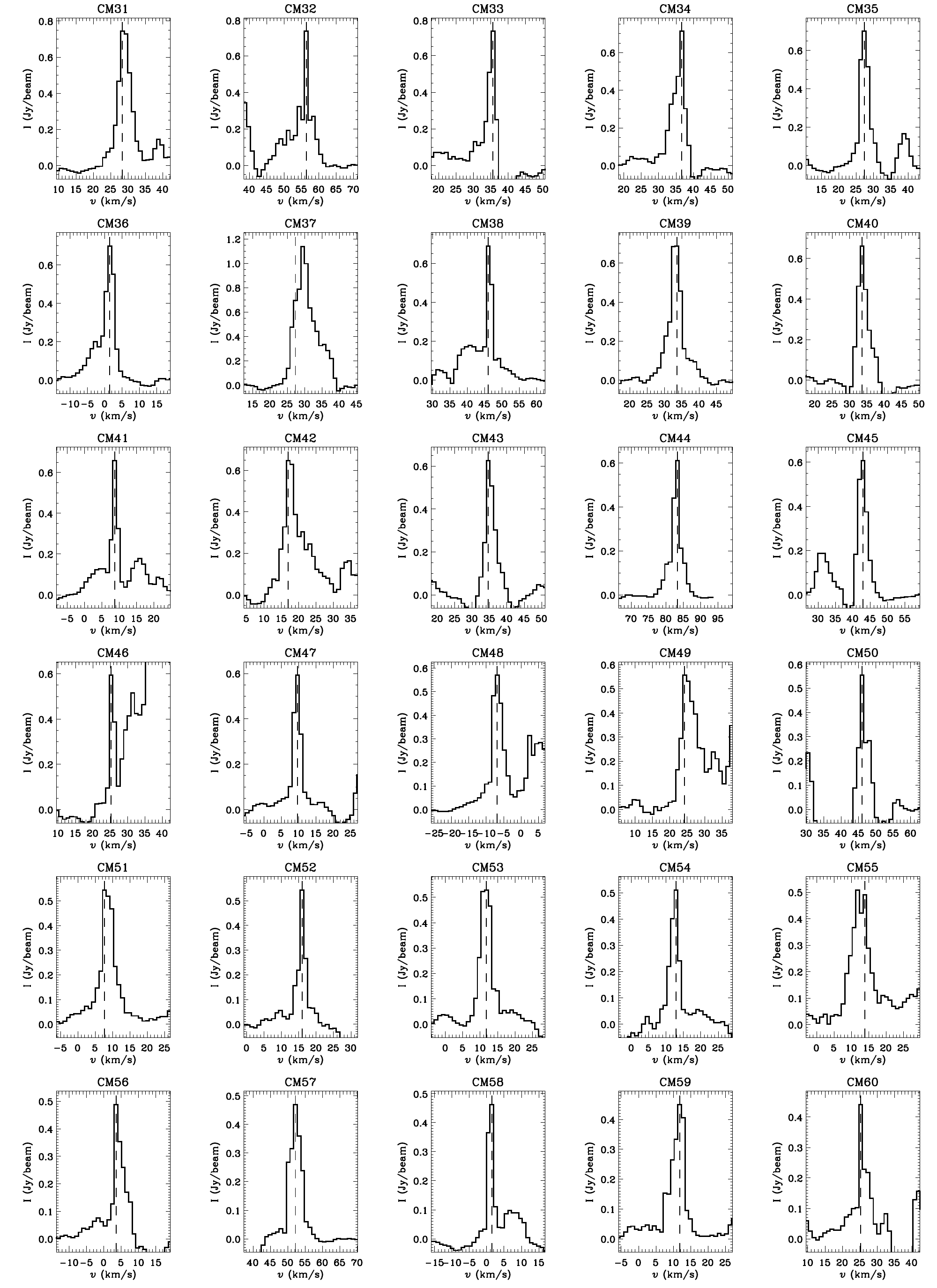}
\caption{ Candidate Maser spectra, continued.}
\label{candspect2}
\end{figure}

\renewcommand\thefigure{8}
\begin{figure}
\includegraphics[scale=0.85]{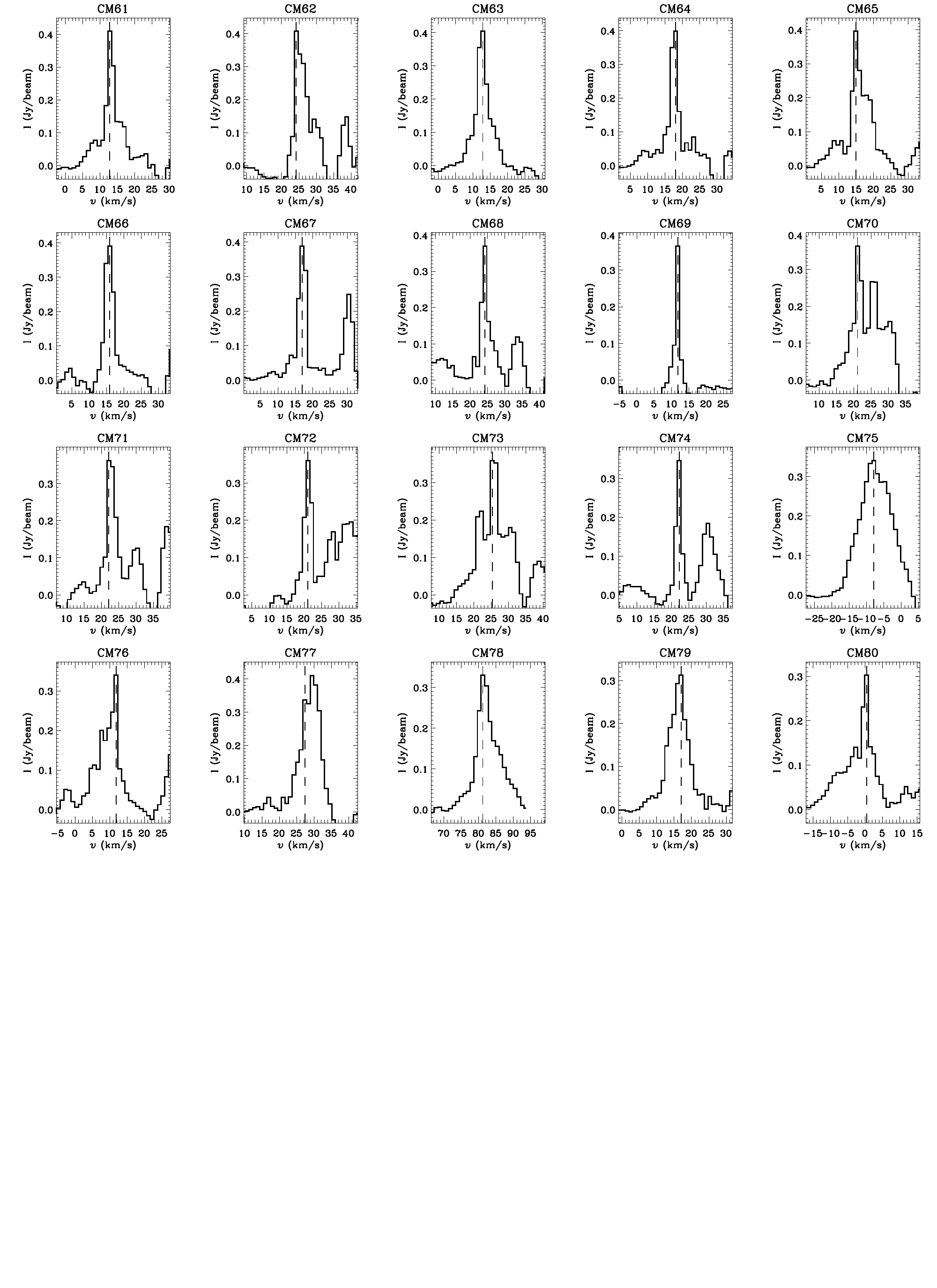}
\caption{ Candidate Maser spectra, continued.}
\label{candspect3}
\end{figure}

\renewcommand\thefigure{9}
\begin{figure*}
\includegraphics[scale=0.75]{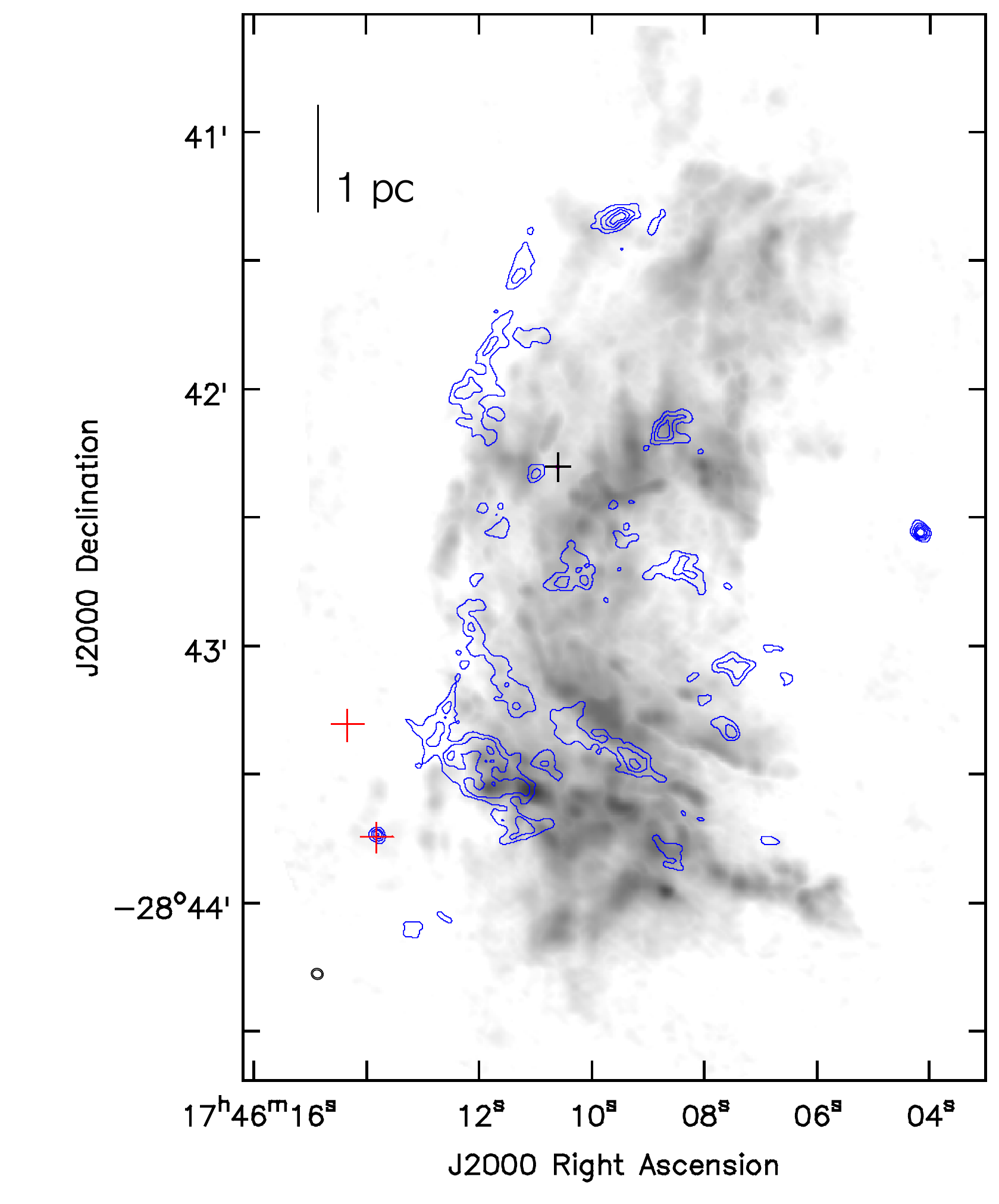}
\caption{ Comparison of the K-band continuum (contours) and the molecular gas, traced by \am\, (3,3). The black cross indicates the location of the H$_2$O maser from \cite{Lis94}, while the red crosses are candidate YSOs from \cite{An11}}
\label{cont_cont}
\end{figure*}

\end{document}